\documentclass[10pt,journal,compsoc]{IEEEtran}

\usepackage{times}
\usepackage{epsfig}
\usepackage{graphicx}
\usepackage{amsmath}
\usepackage{amssymb}
\usepackage{colortbl}
\usepackage{color}
\usepackage{xcolor}
\usepackage{csquotes}
\usepackage{multirow}
\usepackage{multicol}
\usepackage[export]{adjustbox}
\usepackage{xspace}
\usepackage{soul}
\usepackage{gensymb}
\usepackage{csquotes}

\newcommand{\latinphrase}[1]{\textit{#1}} 
\newcommand{\etal}{\latinphrase{et~al.}\xspace}
\newcommand{\ie}{\latinphrase{i.e.}\xspace}

\newcommand{\eg}{\latinphrase{e.g.}\xspace}
\newcommand{\etc}{\latinphrase{etc.}\xspace}

\newcommand{\ch}{\checkmark}

\ifCLASSOPTIONcompsoc
  \usepackage[nocompress]{cite}
\else
  \usepackage{cite}
\fi

\ifCLASSINFOpdf
\else
\fi

\begin{document}
\title{Densely Residual Laplacian Super-Resolution}
\author{Saeed~Anwar,~\IEEEmembership{Member,~IEEE,}
        and~Nick~Barnes,~\IEEEmembership{Senior~Member,~IEEE}
\IEEEcompsocitemizethanks{\IEEEcompsocthanksitem Saeed Anwar is a research fellow with Data61-CSIRO.\protect\\
E-mail: saeed.anwar@data61.csiro.au
\IEEEcompsocthanksitem Nick Barnes is team lead at CSIRO and Associate Professor at ANU.}%
}

\IEEEtitleabstractindextext{%
\begin{abstract}
Super-Resolution convolutional neural networks have recently demonstrated high-quality restoration for single images. However, existing algorithms often require very deep architectures and long training times. Furthermore, current convolutional neural networks for super-resolution are unable to exploit features at multiple scales and weigh them equally, limiting their learning capability. In this exposition, we present a compact and accurate super-resolution algorithm namely, Densely Residual Laplacian Network (DRLN). The proposed network employs cascading residual on the residual structure to allow the flow of low-frequency information to focus on learning high and mid-level features. In addition, deep supervision is achieved via the densely concatenated residual blocks settings, which also helps in learning from high-level complex features. Moreover, we propose Laplacian attention to model the crucial features to learn the inter and intra-level dependencies between the feature maps. Furthermore, comprehensive quantitative and qualitative evaluations on low-resolution, noisy low-resolution, and real historical image benchmark datasets illustrate that our DRLN algorithm performs favorably against the state-of-the-art methods visually and accurately.      
\end{abstract}

\begin{IEEEkeywords}
Super-resolution,  Laplacian attention, Multi-scale attention, Densely connected residual blocks, Deep convolutional neural network.
\end{IEEEkeywords}}
\maketitle

\IEEEdisplaynontitleabstractindextext
\IEEEpeerreviewmaketitle
\IEEEraisesectionheading{\section{Introduction}
\label{sec:introduction}}
\vspace*{-2mm}
\IEEEPARstart{I}{n} recent years, super-resolution (SR), a low-level vision task, became a research focus due to the high demand for better-resolution image quality. Super-resolution addresses the problem of reconstructing a high-resolution (HR) input from a low-resolution (LR) counterpart. We aim to super-resolve a single low-resolution image, a technique, commonly, known as single image super-resolution (SISR). Image SR is a challenging task to achieve as the process is ill-posed, which means that mapping between the output HR image to the input LR image is many-to-one. However, despite being a difficult problem, it is useful in many computer vision applications such as surveillance imaging~\cite{mudunuri2016low}, medical imaging~\cite{greenspan2008super}, forensics \cite{swaminathan2008digital}, object classification~\cite{he2016ResNet} \etc


\begin{figure*}
\begin{center}
\begin{tabular}{c@{ } c@{ }  c@{ } c@{ } c@{ } c@{ } c}
    
    \includegraphics[width=.14\textwidth,valign=t]{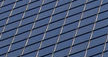}&
    \includegraphics[width=.14\textwidth,valign=t]{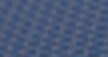}&
    \includegraphics[width=.14\textwidth,valign=t]{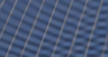}&    
    \includegraphics[width=.14\textwidth,valign=t]{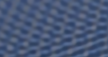}&
    \includegraphics[width=.14\textwidth,valign=t]{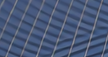}&
    \includegraphics[width=.14\textwidth,valign=t]{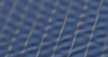}&
    \includegraphics[width=.14\textwidth,valign=t]{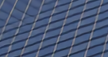}\\
    Ground-truth & Bicubic     &EDSR~\cite{lim2017EDSR} & MSLapSRN~\cite{MSLapSRN}   & RCAN~\cite{zhang2018RCAN}    &CARN~\cite{ahn2018CARN}    & Ours\\
    
    \includegraphics[width=.14\textwidth,valign=t]{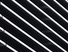}&
    \includegraphics[width=.14\textwidth,valign=t]{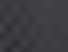}&
    \includegraphics[width=.14\textwidth,valign=t]{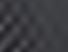}&
    \includegraphics[width=.14\textwidth,valign=t]{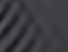}&
    \includegraphics[width=.14\textwidth,valign=t]{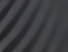}&
    \includegraphics[width=.14\textwidth,valign=t]{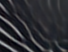}&
    \includegraphics[width=.14\textwidth,valign=t]{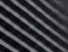}\\
    Ground-truth & Bicubic     &VDSR~\cite{kim2016VDSR} & LapSRN~\cite{lai2017LapSRN} &MSLapSRN~\cite{MSLapSRN}   & RCAN~\cite{zhang2018RCAN}       & Ours\\

\end{tabular}
\end{center}
\vspace*{-4mm}
\caption{\textbf{Visual Comparisons.} Sample results on URBAN100 with Bicubic (BI) degradation for 4$\times$ on \enquote{img\_074} and for  8$\times$ on \enquote{img\_040}. Our method recovers the structures correctly with less distortion and more faithful to the ground-truth image.}
\label{fig:crop_4x_8x}
\vspace*{-5mm}
\end{figure*}

Deep convolutional neural network (CNN) super-resolution methods~\cite{dong2016SRCNNPAMI,zhang2018RCAN,tai2017DRRN} have shown improvement over traditional super-resolution methods in 
SISR. The performance and depth of convolutional neural super-resolution networks have evolved dramatically in recent times. As an example, SRCNN~\cite{dong2016SRCNNPAMI} has three convolutional layers while RCAN~\cite{zhang2018RCAN} has more than 400.
However, using deep networks may be unsuitable for many applications. In this regard, it is essential to design efficient networks.
The most straightforward way to reduce the size of the network is simply to reduce the depth, but this will decrease the quality. Therefore, it is essential to design an efficient network that focuses on reusability of the computed features. 

An effective alternative to depth reduction is to employ recursive architectures, and such attempts are formulated in the form of DRCN~\cite{kim2016DRCN} and DRRN~\cite{tai2017DRRN}. DRCN~\cite{kim2016DRCN} avoids redundant parameters via recursive connections while DRRN~\cite{tai2017DRRN} share's parameters through residual recursive connections. The recursive nets achieved a decrease in the number of parameters, and an increase in performance compared to standard CNN's; however, these models have some limitations, which are: 1) the upsampled input, 2) increased depth and 3) increased width. Although these enable the model to reconstruct the structural features from the low-resolution image, it is at the cost of a large number of operations and high inference time. Another approach to forming a compact model is to utilize the dense connections between convolutional layers~\eg SRDenseNet~\cite{tong2017image} and RDN~\cite{zhang2018RDN}.   

To optimize speed and the number of parameters CARN~\cite{ahn2018CARN} employed group convolutions. The network is primarily based on a variant of residual blocks. Although it can achieve good speed and fewer parameters, it failed to reach the PSNR standard set by RCAN~\cite{zhang2018RCAN}. On the other hand, most of the CNN models~\cite{lim2017EDSR,ledig2017SRresNet,kim2016VDSR,kim2018ram} treat features equally or only at one scale, and therefore, lack adaptability to deal with various frequency levels, \eg low, mid and high. Super-resolution algorithms aim to restore mid-level and high-level frequencies as the low-level frequencies can be obtained from the input low-resolution image without substantial computations. The state-of-the-art methods~\cite{zhang2018RCAN,zhang2018RDN,haris2018DDBPN}, models the features equally or on a limited scale, ignoring the abundant rich frequency representation at other scales; hence these lack discriminative learning capability and capacity across channels, and eventually, this limits the ability of convolutional neural networks. To address these issues, we propose the densely residual Laplacian attention Network (DRLN) to reconstruct SR images. DRLN utilizes the dense connection between the residual blocks to use the previously computed features. Similarly, we employ Laplacian pyramid attention to weight the features at multiple scales and according to their importance.   

In summary, our main contributions are four-fold: 
\vspace*{-3mm} 
\begin{itemize} 
\item We propose densely connected residual blocks and a Laplacian attention network for accurate image super-resolution. Our network achieves much better performance through multi-shortcut connections and multi-level representation. 

\item Our novel design employs cascading over residual on the residual architecture, which can assist in training deep networks. Diverse connection types and cascading over residual on the residual in our DRLN help in bypassing enough low-frequency information to learn more accurate representations. 

\item We introduce Laplacian attention, which has a two-fold purpose: 1) To learn the features at multiple sub-band frequencies and 2) to adaptively rescale features and model feature dependencies. Laplacian attention further improves the feature capturing capability of our network. 


\item Through extensive experiments, we show DRLN is efficient and achieves better performance.
\vspace*{-5mm} 
\end{itemize}

\section{Related Works}
\label{sec:related_works}
In this section of the paper, we provide chronological advancement in the deep super-resolution. Dong~\etal~\cite{dong2016SRCNNPAMI} proposed pioneering works in super-resolution by introducing a fully convolutional network composed of three convolutional layers followed by ReLU~\cite{he2015PReLU} and termed it as SRCNN~\cite{dong2016SRCNNPAMI}. The input to the SRCNN~\cite{dong2016SRCNNPAMI} is a bicubic interpolated image which diminishes high-frequencies and requires additional computation. To reduce the burden on the network, FSRCNN~\cite{dong2016FSRCNN} inputs the original low-resolution image and employ deconvolution to upsample the features to the desired dimensions before the final objective function. The authors of~\cite{dong2016FSRCNN} also uses the shrinking and expansion of channels to make the model near real-time on a CPU.  

Initially, the focus was on linear networks, bearing a simple architecture with no skip-connections \ie only one path for the signal flow with the layers stacked consecutively. SRCNN~\cite{dong2016SRCNNPAMI} and FSRCNN~\cite{dong2016FSRCNN} are examples of linear networks. Similarly, Image Restoration CNN abbreviated as IRCNN~\cite{zhang2017IrCNN}, another straight model, can restore several low-level vision tasks jointly. The aim here is to employ dilation in convolutional layers to capture a larger receptive field for better learning coupled with batch normalization and non-linear activation (ReLU) to reduce the depth of the network. Furthermore, SRMD~\cite{zhang2018SRMDNF}, an extended super-resolution network, can handle different degradations. SRMD~\cite{zhang2018SRMDNF} inputs low-resolution images and their computed degradation maps. The model structure is similar to~\cite{dong2016SRCNNPAMI,zhang2017IrCNN}.  

With the emergence of skip-connections in CNN networks, its usage became a prominent feature in super-resolution. In this regard, very deep super-resolution (VDSR)~\cite{kim2016VDSR} incorporated a global skip connection to enforce residual learning using gradient clipping to avoid gradient vanishing. VDSR~\cite{kim2016VDSR} improved upon the previous CNN super-resolution methods. Inspired from VDSR~\cite{kim2016VDSR}, the same authors next presented DRCN~\cite{kim2016DRCN}, which shares parameters using a deep recursive structure. This sharing technique reduces the number of parameters significantly; however, the performance is lagging behind VDSR~\cite{kim2016VDSR}. Subsequently, deep recursive residual network (DRRN)~\cite{tai2017DRRN} replicates primary skip-connections across different convolutional blocks to enforce residual learning through multi-path architecture. The aim is to reduce the memory cost and computational complexity via parameter sharing. Further, Tai~\etal~\cite{tai2017memnet} introduces a persistent memory network (MemNet), which is composed of memory blocks stacked together recursively. Each block is then connected to a gate unit densely, where each gate unit is a convolutional layer with kernel size 1$\times$1. The performance of the networks employing recursive connections is comparable to each other.  

Lim~\etal~\cite{lim2017EDSR} proposed the enhanced deep super-resolution (EDSR) network, which employs residual blocks and a long skip-connection. EDSR~\cite{lim2017EDSR} rescaled the features by a factor of 0.1 to avoid gradient exploding. EDSR improved upon all previous methods by a significant margin. More recently, Ahn~\etal~\cite{ahn2018CARN} proposed the cascading residual network (CARN) which also employs a variant of residual blocks \ie having three convolutional layers as compared to the customarily-used two convolutional layers with cascading connections. CARN~\cite{ahn2018CARN} lags behind EDSR~\cite{lim2017EDSR} in terms of PSNR.   

Driven by the success of the dense-connection architecture proposed in DenseNet~\cite{huang2017densely} by Huang~\etal for image classification, super-resolution networks have focused on the dense-connections to improve performance. As an example, SRDenseNet~\cite{tong2017image} utilized dense-connections where every convolutional layer in a block operates on the output of all prior convolutional layers. To upsample the features, SRDenseNet~\cite{tong2017image} orders the blocks sequentially followed by deconvolutional layers at the end of the network. Likewise, Zhang~\etal~\cite{zhang2018RDN} proposed a residual dense network (RDN) to learn local features from the images via dense-connections. Furthermore, to avoid vanishing gradients and for ease of flow of information from low-level to high-level layers, RDN~\cite{zhang2018RDN} employed skip-connections. Lately, DDBPN~\cite{haris2018DDBPN} aims to model a feedback mechanism with a feed-forward procedure; hence, a series of densely connected upsampling and downsampling layers are used as a single block. To predict the final super-resolved image, the outputs of the intermediate blocks are concatenated as well.   

\begin{figure*}[t]
\begin{center}
\includegraphics[trim={1cm 3cm 0.5cm, 1.5cm},clip,width=1\textwidth]{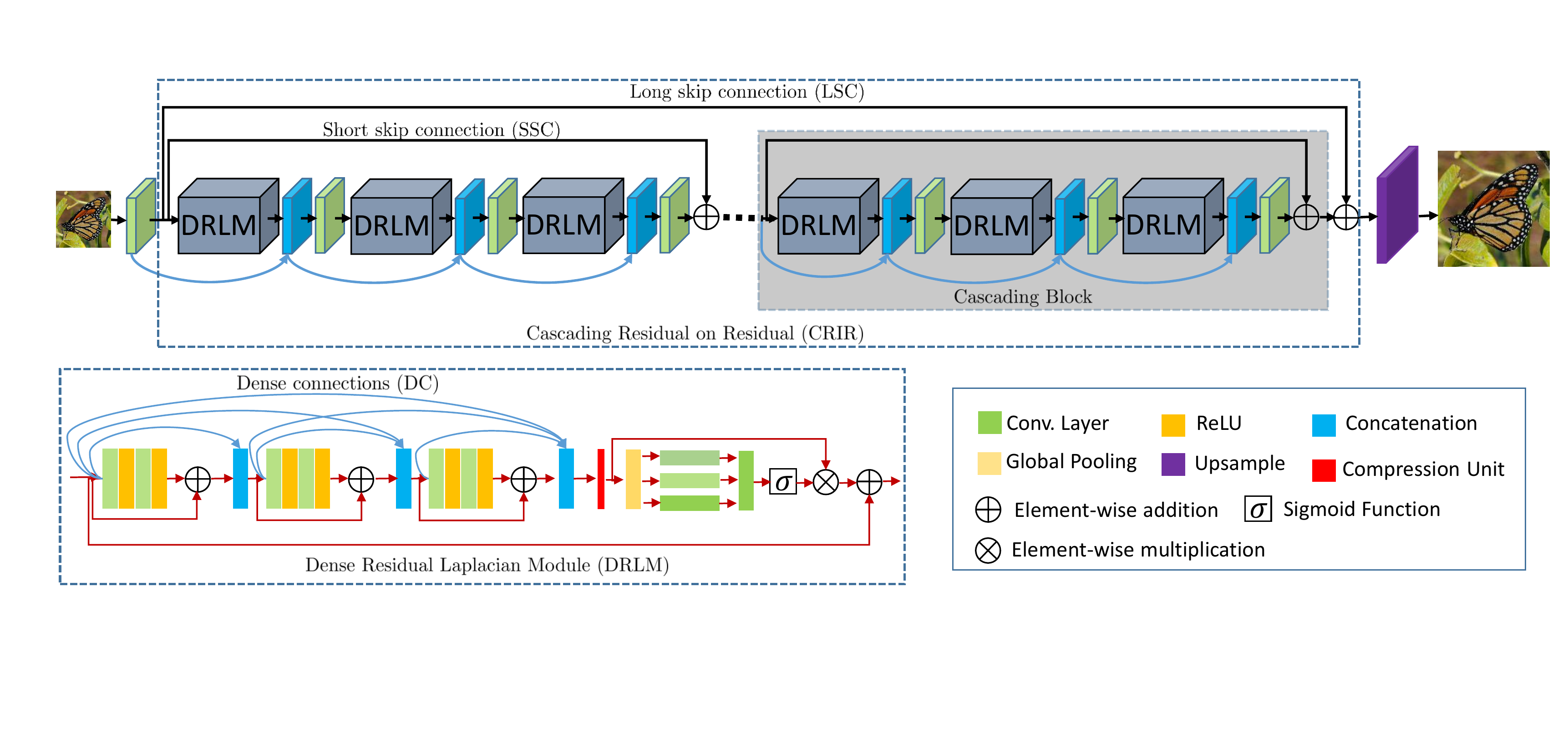}
\end{center}
\vspace*{-4mm}
\caption {\textbf{The detailed network architecture of the proposed Network.} The top figure shows the overall architecture of our proposed network with cascading residual on the residual architecture \ie a long skip connection, short skip connections, and cascading structures. The bottom figure presents the backbone of our network \ie Dense Residual Laplacian Module (DRLM).}
\label{fig:net}
\vspace*{-5mm}
\end{figure*}

To obtain distinct features at multiple scales, multi-branch networks~\cite{Ren2017CNF,hu2018CMSC,hui2018fast} are proposed. Ren~\etal~\cite{Ren2017CNF} employ SRCNN~\cite{dong2014SRCNN} at various branches with a different number of layers to learn features uniquely and lastly combine them using a sum-pooling layer. Similarly, Hu~\etal~\cite{hu2018CMSC} proposed cascaded multi-scale cross-network composed of subnets. Each subnet has merge-and-run units consisting of two parallel branches, each having two convolutional layers. Batch normalization and Leaky-ReLU \cite{maas2013LeakyReLU} follows each convolutional layer in the merge-and-run unit. In contrast to multi-branch, Lai~\etal~\cite{lai2017LapSRN} proposed a multi-stage network where each sub-network progressively predicts the residual output up to an 8$\times$ factor.  

To enhance the visual quality of the images, Generative Adversarial Networks (GANs)~\cite{radford2015supervisedGAN,goodfellow2014generative} aim to improve the perceptual quality through super-resolution. The first exciting work in this regard is SRResNet~\cite{ledig2017SRresNet}, where the generator is comprised of residual blocks similar to~\cite{he2016deep} with a skip-connection from the input to the output while the discriminator is fully convolutional. The SRResNet~\cite{ledig2017SRresNet} combined three different losses, which include perceptual, adversarial and $\ell_2$. Next, to create the textures faithful to the original image, EnhanceNet~\cite{sajjadi2017enhancenet} used an additional texture matching loss with the mentioned losses. This loss aims to match the textures of low-resolution and high-resolution patches as gram matrices computed from deep features via the $\ell_1$.   

Similar to~\cite{sajjadi2017enhancenet}, to generate more realistic super-resolved images, Park~\etal~\cite{park2018srfeat} proposed SRFeat, which utilizes an additional discriminator to help the generator. The results of SRFeat~\cite{park2018srfeat} are perceptually better than~\cite{sajjadi2017enhancenet}. Inspired by~\cite{ledig2017SRresNet} network, ESRGAN~\cite{wang2018esrgan} removed the batch normalization and used dense-connections between the convolutional layers in the same segment. A global skip-connection is incorporated for residual learning. Besides, changing the elements of the generator, an enhanced discriminator \ie Relativistic GAN~\cite{jolicoeur2018relativisticGAN} is used instead of the traditional one. The performance of the ESRGAN~\cite{wang2018esrgan} is the best among the current super-resolution GAN algorithms. Furthermore, the GAN super-resolution models have significantly improved the perceived quality compared to its CNN competitors.  

Visual attention~\cite{mnih2014recurrent} is primarily employed in image classification. This concept was brought to image super-resolution by RCAN~\cite{zhang2018RCAN}, which uses a channel attention mechanism for modeling the inter-channel dependencies coupled with stacking of groups of residual blocks. The PSNR values of RCAN~\cite{zhang2018RCAN} is the best among all the algorithms as mentioned earlier. In parallel to RCAN~\cite{zhang2018RCAN}, Kim~\etal~\cite{kim2018ram} proposed a dual attention mechanism, namely, the super-resolution residual attention module (SRRAM). The depth of the SRRAM~\cite{kim2018ram} is comparatively smaller than RCAN~\cite{zhang2018RCAN} and lag behind RCAN~\cite{zhang2018RCAN} in PSNR numbers. On the other hand, our method improves upon RCAN~\cite{zhang2018RCAN} both visually and in numbers by exploiting densely connected residual blocks followed by multi-scale attention using different levels of the skip and the cascading connections.

\begin{figure*}[t]
\begin{center}
\includegraphics[trim={3.5cm 6.5cm 1.5cm, 3cm},clip,width=1\textwidth]{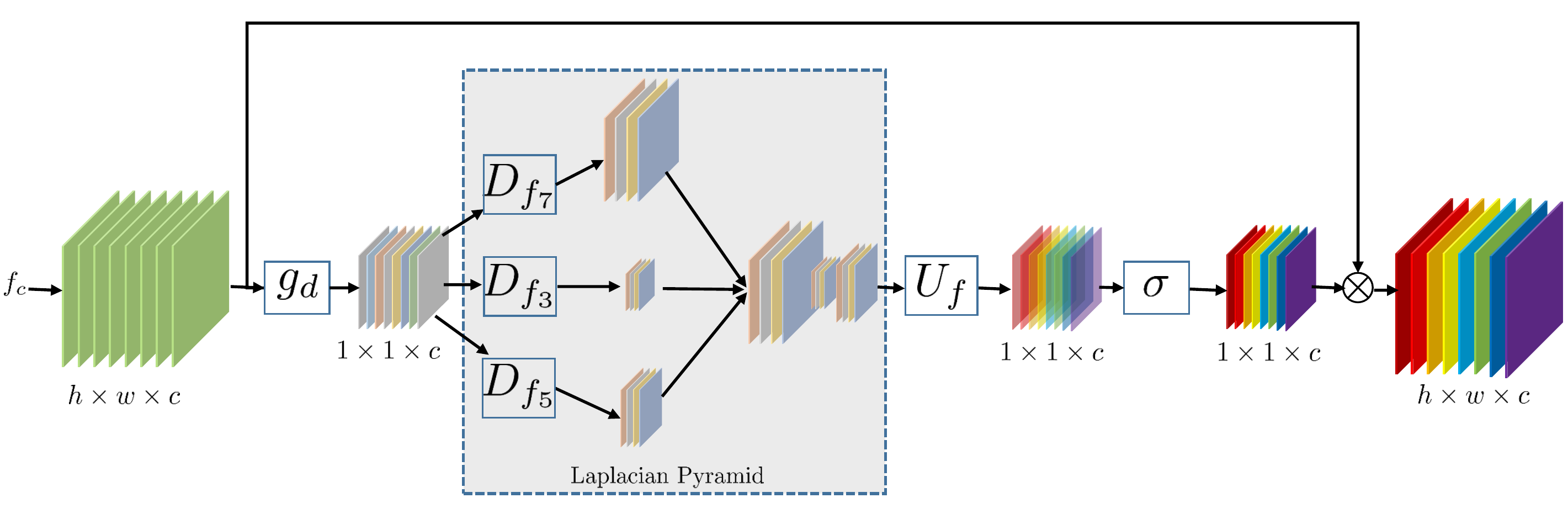}
\end{center}
\vspace*{-4mm}
\caption{\textbf{Laplacian attention.} Our model consists of pyramid-level attention to model the features non-linearly. The Laplacian attention weights the residual features at different sub-frequency-bands.}
\label{fig:laplacian_att}
\vspace*{-4mm}
\end{figure*}

\section{Our Model}
\subsection{Network Architecture}
Our model is constituted of four integral components, \ie, \emph{feature extraction}, \emph{cascading over residual on the residual}, \emph{upsampling}, and \emph{reconstruction}, as shown in Figure~\ref{fig:net}. Let's suppose the low-resolution input image, and the super-resolved output image is represented by $x$ and $\hat{y}$, respectively. To formally illustrate the model implementation, let $f$ be a convolutional layer and $\tau$ be a non-linear activation function; then, we define the feature extraction component which is comprised of one convolutional layer to extract primitive features from the low-resolution input, as:

\begin{equation} 
f_0 = H_f(x), 
\label{eq:extraction} 
\end{equation} 
where $H_f(\cdot)$ is the convolutional operator applied on the low-resolution image. Next, $f_0$ is passed on to the cascading residual on the residual component, termed as $H_{crir}$, 
\begin{equation} 
f_r = H_{crir}(f_0), 
\label{eq:drs} 
\end{equation} 
where $f_r$ are the estimated features and $H_{crir}(\cdot)$ is the main cascading residual on the residual component which is composed of dense residual Laplacian modules cascaded together. The output features of the $H_{crir}$ are novel to the best of our knowledge in image super-resolution. Our method's depth is not significant compared to RCAN~\cite{zhang2018RCAN}; however, it provides a wide receptive field and the best results. Following this, the extracted deep $f_r$ features from the cascaded residual on the residual component are upscaled through the upsampling component, as:  

\begin{equation} 
f_u = H_u(f_r), 
\label{eq:upsample} 
\end{equation} 
where $H_u(\cdot)$ and $f_u$ denote an upsampling operator and upscaled features, respectively. Although several choices are available for $H_u(\cdot)$ such as a deconvolutional layer~\cite{dong2016FSRCNN}, or nearest-neighbor upsampling with convolution~\cite{dumoulin2017learned}; we opt for ESPCN~\cite{shi2016ESPCN} following the footsteps of~\cite{zhang2018RCAN,lim2017EDSR}. Next, the $f_u$ features are passed through the reconstruction component which is composed of one convolutional layer to predict the super-resolved RGB color channels as an output, expressed as:

\begin{equation} 
\hat{y} = H_r(f_u), 
\label{eq:r} 
\end{equation} 
where $\hat{y}$ is the estimated super-resolved image while $H_r(\cdot)$ denotes the reconstruction operator. 

To optimize our model, several choices are available for the loss function, including $\ell_2$~\cite{dong2016SRCNNPAMI,kim2016VDSR}, $\ell_1$~\cite{lim2017EDSR,zhang2018RCAN}, perceptual~\cite{sajjadi2017enhancenet}, total variation~\cite{jiao2017formresnet} and adversarial~\cite{wang2018esrgan,ledig2017SRresNet}. To be fair with the competing state-of-the-art methods~\cite{lim2017EDSR,zhang2018RDN,zhang2018RCAN}, we also choose $\ell_1$ loss function for our network optimization. Now, for a batch of $N$ training pairs, \ie $\{x_i, y_i\}_{i=1}^N$, the aim is to minimize the $\ell_1$ loss function as 
\begin{equation} 
L(\mathcal{W}) = \frac{1}{N} \sum_{i=1}^N||\text{DRLN}(x_i) - y_i||_1, 
\label{eq:l1_loss} 
\end{equation} 
where DRLN($\cdot$) is our network and $\mathcal{W}$ denotes the set of all the network parameters learned. The feature extraction $H_f$ and reconstruction $H_r$ are similar to previous algorithms~\cite{kim2016VDSR,dong2016FSRCNN}. In the next section, we focus on the $H_{crir}$.

\subsection{Cascading Residual on the Residual}
\label{ss:RIR}
In this section, we provide more details on the cascading residual on the residual structure, which has a hierarchical architecture and composed of \emph{cascading blocks}. Each cascading block has a medium skip-connection (MSC), cascading features concatenation and is made up of \emph{dense residual Laplacian modules (DRLM)} each of which consists of a \emph{densely connected residual unit}, \emph{compression unit} and \emph{Laplacian pyramid attention unit}. The lower part of Figure~\ref{fig:net} shows DRLM, which will be explained further in Section~\ref{ss:BM}.

Recently, in image recognition~\cite{he2016deep} and super-resolution~\cite{zhang2018RCAN}, residual blocks are stacked together to construct a network of more than 1000 layers and 400 layers, respectively, via skip-connections, although the performance has increased; however, the computational overhead has also increased. We aim here to construct a compact and efficient model with a much lower number of the convolutional layers amidst improved performance and computation time. Therefore, we introduce cascading of the blocks employing medium and long skip connections. Let's suppose that the $n$-th dense residual Laplacian module (DRLM) of the $m$-th cascaded block $B^{n,m}$ is given as: 
\begin{equation} 
f_{n,m} = f([Z^{u-0;m}, Z^{u-1;m}, B^{u;m}(w^{u,1;m}),b^{u,1;m} ) 
\label{eq:rir_block} 
\end{equation} 
where $f_{n,m}$ are the features from the $n$-th dense residual Laplacian module (DRLM) of the $m$-th cascaded block. Each cascaded block is composed of $k$ DRLMs, and hence the input to the cascaded block is summed with the output of the $k$ DRLM as $f_{n+k,m} = f_{n+k,m}+f_{n,m}$, \ie medium skip-connection (MSC) as:  

\begin{equation} 
f_g = f_0 + H_{crir}(\mathcal{W}_{w,b}), 
\label{eq:residual_block} 
\end{equation} 
where $\mathcal{W}_{w,b}$ are the weights and biases learned in the cascaded block. The addition of medium skip-connection eases the flow of information across group of DRLM while the addition of long-skip connection (LSC) helps the flow of information through cascaded blocks. The group features $f_g$ are passed to the reconstruction layer to output the same number of channels as the input to the network. Next, we provide information about the dense residual Laplacian modules and its subcomponents.   

\subsection{Dense Residual Laplacian Module}
\label{ss:BM}

As briefly mentioned earlier, DRLM is composed of three subcomponents \ie \emph{densely connected residual blocks unit}, \emph{compression unit}, and \emph{Laplacian pyramid attention}.  

The residual blocks we employ, have the traditional two convolutional layers and two ReLUs structure followed by an addition from the input of the residual block, as: 
\vspace*{-0.25mm}
\begin{equation} 
R_{i}(w_i,b_i) = \tau(f(\tau(f(w_{i,1},b_{i,1});b_{i,2})+Z_{i-1}), 
\label{eq:RB_block} 
\end{equation} 
where $Z_{i-1}$ is the output of the previous convolutional layer or residual block while $w_{i,j}$ are the weights of the convolutional layer and $b_{i,j}$ are the biases ($j \in \{1,2\}$). Each densely connected residual unit has three residual blocks with dense connection as  
\vspace*{-0.25mm}
\begin{equation} 
\begin{split} 
R_c &= [R_{i-1}(w_{i-1},b_{i-1});R_{i-2}(w_{i-2},b_{i-2})],\\ 
f_R &= R_{i}(w_{i},b_{i}) = \tau(f(\tau(f(R_c))+R_c), 
\end{split} 
\label{eq:DC_block} 
\end{equation} 
where $R_c$ is the concatenation of the previous residual blocks and $f_R$ is the final output of the densely connected residual unit. The $f_R$ features are then passed through a compression unit which compresses the high number of parameters resulted from dense concatenation. The compression unit is comprised of a single convolutional layer with a kernel of 1$\times$1. The compressed features $f_c$ are then forwarded to the Laplacian attention unit which is described next. 


\textbf{Laplacian Attention:} Image Attention has been employed in image classification~\cite{wang2017residual}, image captioning~\cite{vinyals2015show} \etc to converge on essential image regions. In super-resolution, the same concept with a little variation can be applied that features should be weighted according to their relative importance. Here, we propose Laplacian attention to boost and exploit the relationship between the features that are essential for super-resolving the images. 

To produce attention differently at the Laplacian pyramids in the DRLM, we use a global descriptor to capture the statistics expressing the entire image. The proposed Laplacian pyramid weights the sub-band features of high importance progressively in each DRLM. The global descriptor takes the output from the compression unit \ie $f_c$ which has size $h \times w$ with $c$ feature maps. After processing, the global descriptor reduces the size from $h \times w \times c$ to $1\times1\times c$ as:  


\begin{figure}[t]
\vspace*{-4mm}
\begin{center}
\includegraphics[width=0.48\textwidth]{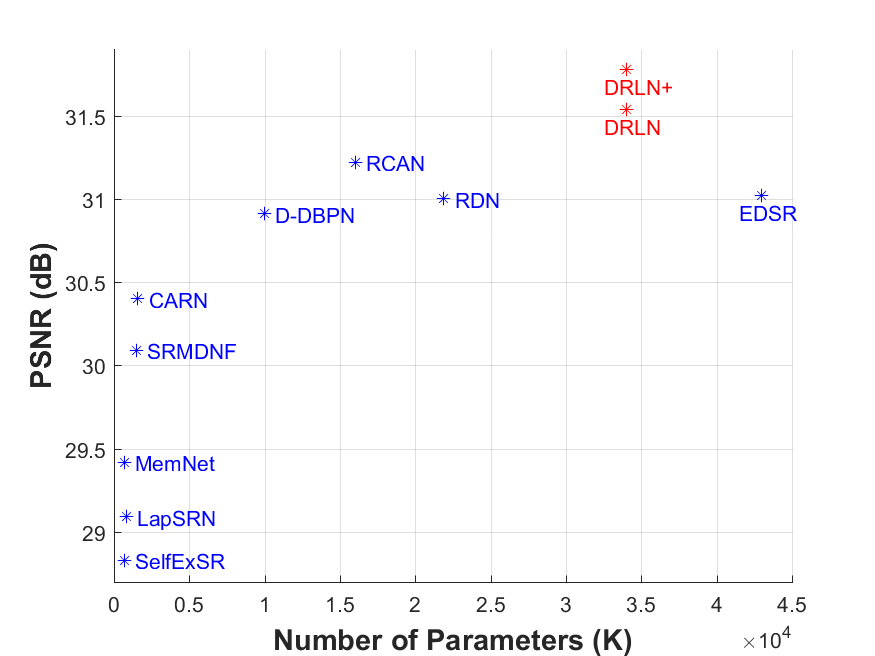}
\vspace*{-4mm}
\caption{\textbf{Parameters vs. performance.} Comparisons are presented on the MANGA109~\cite{fujimoto2016manga109} for 4$\times$ super-resolution.}
\label{fig:PvsP}
\end{center}
\vspace*{-4mm}
\end{figure}

\begin{equation} 
g_d = \frac{1}{h \times w} \sum_{i=1}^h \sum_{i=1}^w f_c(i,j), 
\label{eq:GAP} 
\end{equation} 
where $f_c(i,j)$ is the value at position $(i,j)$ in the feature maps. 

To capture the channel dependencies from the retrieved global descriptor, we utilize a gating approach. As studied in~\cite{hu2017squeeze}, the system must be able to learn the nonlinear synergies between feature maps and mutually-exclusive associations via gating. To implement the gating mechanism formally, we utilize ReLU and sigmoid functions, denoted by $\tau$, and $\sigma$, respectively. The $g_d$ features are passed through the Laplacian pyramid to learn  the critical features at different scales as: 

\begin{equation} 
\begin{split} 
r_3 &= \tau(D_{f_3}(g_d)),\\ 
r_5 &= \tau(D_{f_5}(g_d)),\\ 
r_7 &= \tau(D_{f_7}(g_d)),\\ 
\end{split} 
\label{eq:gating} 
\end{equation} 
where $D\_$ is the feature reduction operator while the $f_3, f_5$ and $f_7$ are the convolutional layers with kernel dilation specified by the subscripts. The multi-level representations $r_3, r_5$ and $r_7$ obtained from the the global descriptor $g_d$ are concatenated as: 

\begin{equation} 
g_p = [r_3; r_5; r_7].\\ 
\label{eq:laplacian_convs} 
\end{equation}

Furthermore, as shown in Eq~\ref{eq:gating}, the output of the Laplacian pyramid is convolved with a downsampling operator. Therefore, to upsample and differentiate between the features maps, the output is then fed into the upsampling operator $U_f$ followed by sigmoid activation as: 
\begin{equation} 
L_p = \sigma(U_f(g_p)). 
\label{eq:sigmoid} 
\end{equation} 

As a final step, the \emph{learned statistics} are utilized by adaptively rescaling the output of sigmoid function \ie $L_p$ by the input $f_c$ of the Laplacian attention unit as: 
\begin{equation} 
\hat{f}_c = L_p\times f_c 
\label{eq:rescale_CA} 
\vspace*{-5mm} 
\end{equation}   
\begin{table*}[t]
\caption{\textbf{Contribution of different components.} Investigation of the performance due to different components of our network.}
\vspace*{-4mm}
\begin{center}
\begin{tabular}{|l||c|c|c|c|c|c|c|c|c|c|c}
\hline \hline
Dense Connections (DC)       &       &\ch      &       &        &      & \ch  & \ch   & \ch   &\ch    &\ch\\ 
Medium Skip Connections (MSC)&       &\ch      & \ch   & \ch    & \ch  &      &       &       &\ch    &\ch\\
Long Skip Connection (LSC)   &       &\ch      & \ch   & \ch    &      & \ch  & \ch   &       &       &\ch\\
Laplacian Attention (LA)     &       &         &       & \ch    &      &      & \ch   &       &       &\ch\\ \hline \hline
PSNR (in dB)                 &31.92  &32.30    & 32.06 & 32.06  &32.07 &31.85 & 32.12 & 31.97 & 32.10 & \textbf{32.37}   \\  \hline
\end{tabular}
\end{center}
\label{table:Skip_connections}
\vspace*{-4mm}
\end{table*}


\begin{figure}[t]
\vspace*{-4mm}
\begin{center}
\includegraphics[width=0.48\textwidth]{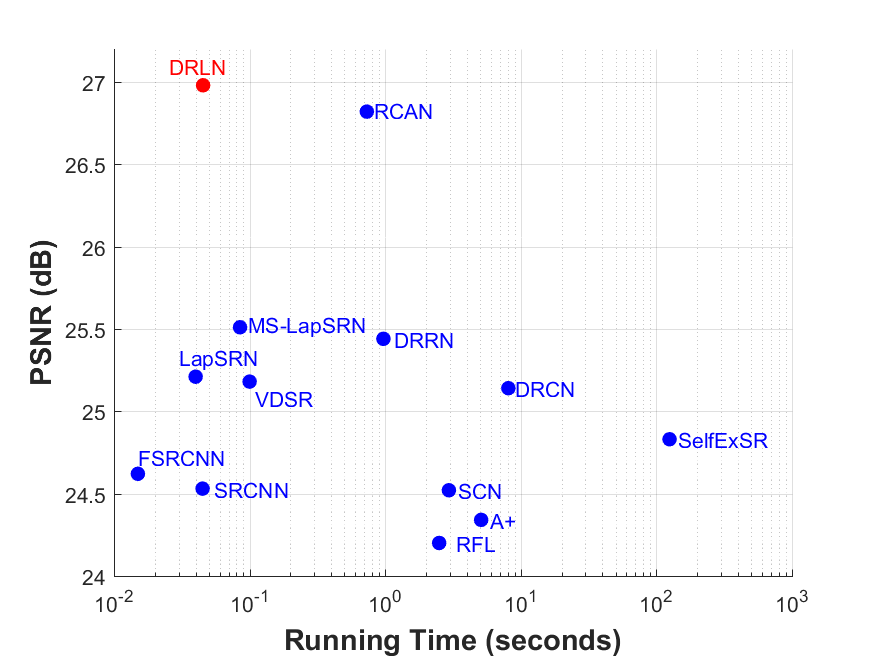}
\vspace*{-4mm}
\caption{\textbf{Performance vs. Time.} Comparisons are presented on the URBAN100~\cite{huang2015URBAN100} for 4$\times$ super-resolution. Our proposed method strides a balance between performance and computation time. }
\label{fig:PvsT}
\end{center}
\vspace*{-4mm}
\end{figure}

\begin{figure*}
\begin{center}
\begin{tabular}{c@{ } c@{ }  c@{ } c@{ } c@{ } c}
    
    \multirow{4}{*}{\includegraphics[width=.24\textwidth,height=2.95cm,valign=t]{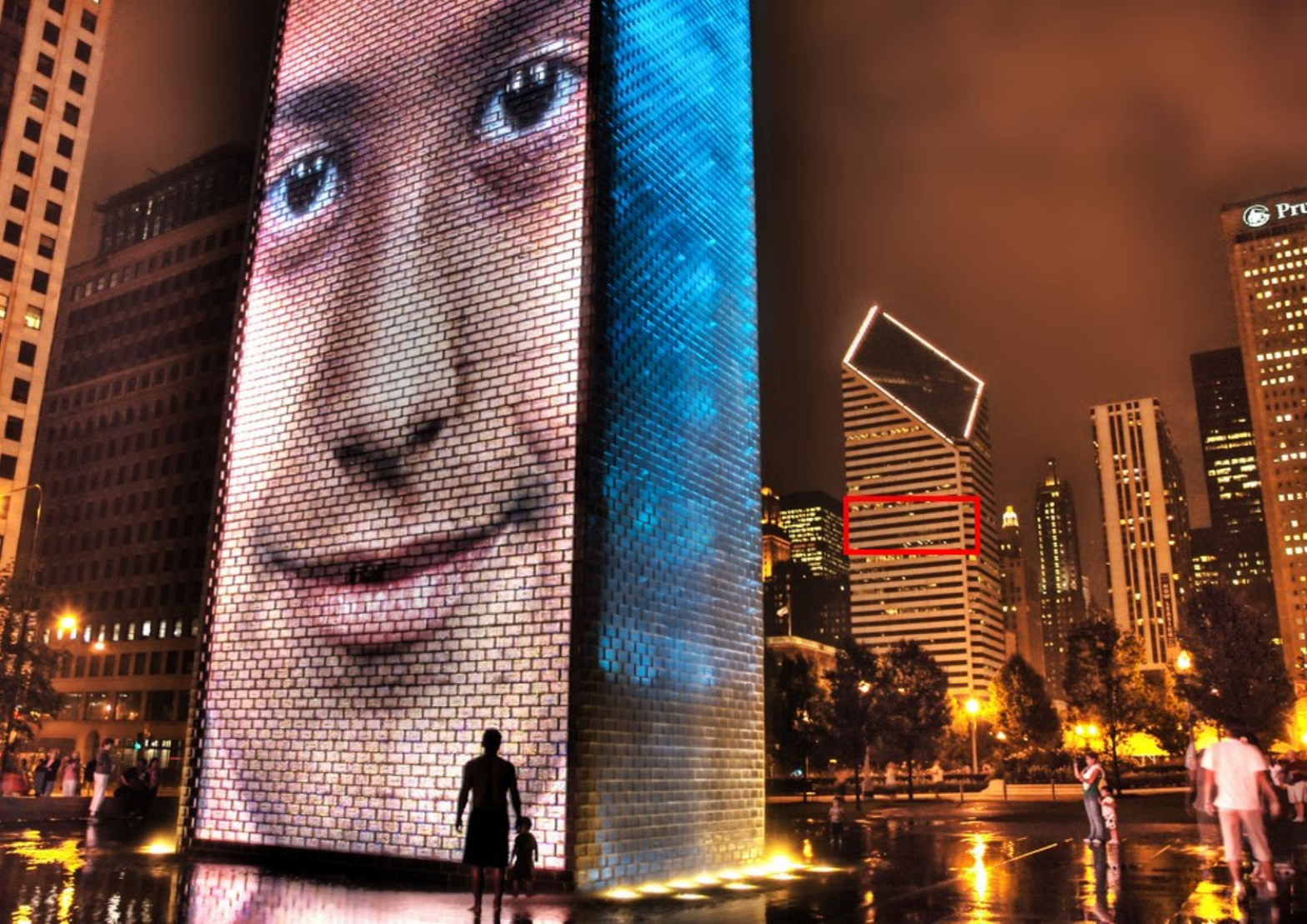}} &  
    \includegraphics[width=.145\textwidth,valign=t]{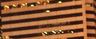}&
  	\includegraphics[width=.145\textwidth,valign=t]{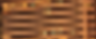}&   
    \includegraphics[width=.145\textwidth,valign=t]{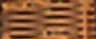}&   
  	\includegraphics[width=.145\textwidth,valign=t]{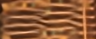}&
  	\includegraphics[width=.145\textwidth,valign=t]{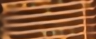}\\
    & Original  & Bicubic       & SRCNN~\cite{dong2016SRCNNPAMI}           & VDSR~\cite{kim2016VDSR}          & MSLapSRN~\cite{MSLapSRN}\\
    &PSNR/SSIM  & 21.58/0.6290  & 22.03/0.6786   & 22.15/0.6925  & 22.31/0.7030\\

    &
     \includegraphics[width=.145\textwidth,valign=t]{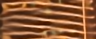}&
    \includegraphics[width=.145\textwidth,valign=t]{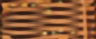}&  
    \includegraphics[width=.145\textwidth,valign=t]{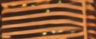}&    
    \includegraphics[width=.145\textwidth,valign=t]{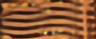}&    \includegraphics[width=.145\textwidth,valign=t]{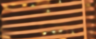}\\
    Urban100 (4$\times$) &DRRN~\cite{tai2017DRRN}          &EDSR~\cite{lim2017EDSR}           & RCAN~\cite{zhang2018RCAN}         & CARN~\cite{ahn2018CARN}         & Ours\\
    img$\_$076           &21.93/0.6903  & 23.07/0.7367  &24.31/0.7897  &22.57/0.7175  &\textbf{24.62/0.8032} \\

    \multirow{4}{*}{\includegraphics[width=.24\textwidth,height=2.8cm, valign=t]{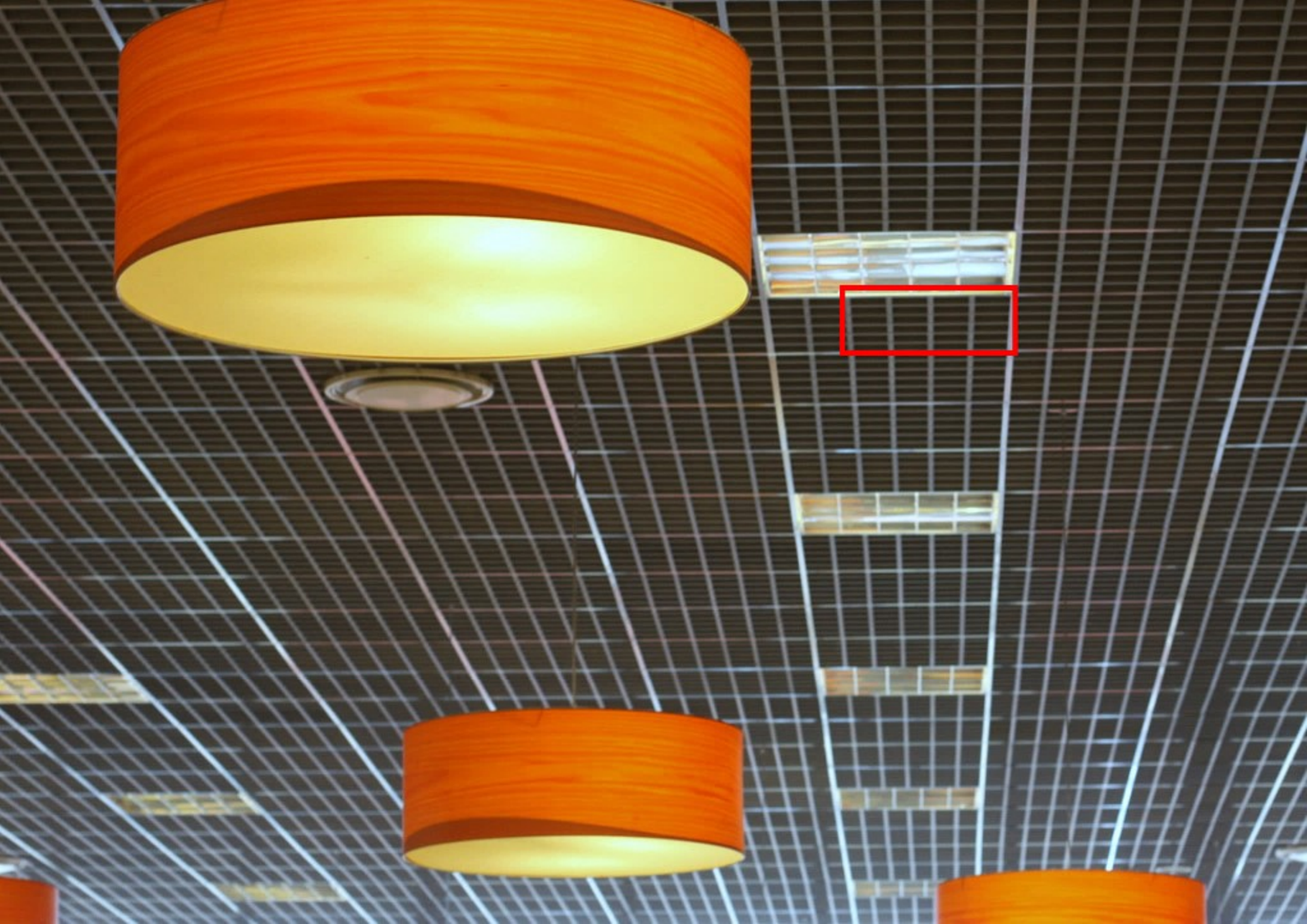}} &  
    \includegraphics[width=.145\textwidth,valign=t]{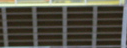}&
  	\includegraphics[width=.145\textwidth,valign=t]{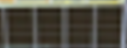}&   
    \includegraphics[width=.145\textwidth,valign=t]{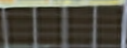}&   
  	\includegraphics[width=.145\textwidth,valign=t]{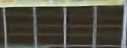}&
  	\includegraphics[width=.145\textwidth,valign=t]{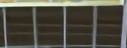}\\
    & Original  & Bicubic       & SRCNN~\cite{dong2016SRCNNPAMI}         & VDSR~\cite{kim2016VDSR}          & MSLapSRN~\cite{MSLapSRN}\\
    &PSNR/SSIM  & 26.92/0.7254  & 29.70/0.8102  & 29.69/0.8312  & 30.03/0.8430 \\

    &
    \includegraphics[width=.145\textwidth,valign=t]{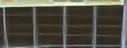}&
    \includegraphics[width=.145\textwidth,valign=t]{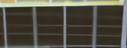}&  
    \includegraphics[width=.145\textwidth,valign=t]{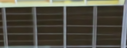}&    
    \includegraphics[width=.145\textwidth,valign=t]{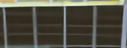}&    \includegraphics[width=.145\textwidth,valign=t]{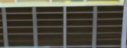}\\
    Urban100 (4$\times$) & DRRN~\cite{tai2017DRRN}           & EDSR~\cite{lim2017EDSR}          & RCAN~\cite{zhang2018RCAN}         & CARN~\cite{ahn2018CARN}          & Ours\\
    img$\_$044           & 29.30/0.8373   & 33.36/0.9054  & 31.45/0.7955 & 31.34/0.8648  & 34.77/0.9188 \\

    \multirow{4}{*}{\includegraphics[width=.24\textwidth,height=3.9cm,valign=t]{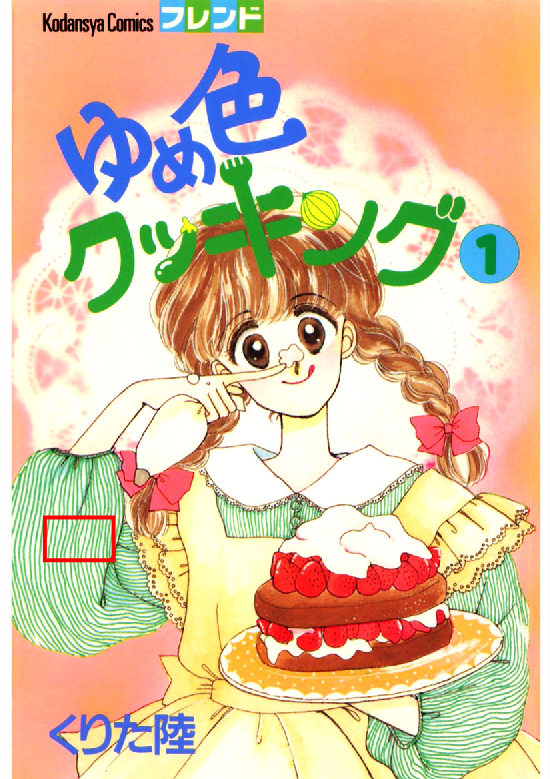}} &  
    \includegraphics[width=.145\textwidth,valign=t]{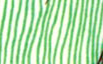}&
  	\includegraphics[width=.145\textwidth,valign=t]{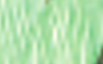}&   
    \includegraphics[width=.145\textwidth,valign=t]{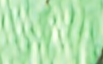}&   
 	\includegraphics[width=.145\textwidth,valign=t]{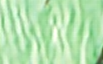}&
  	\includegraphics[width=.145\textwidth,valign=t]{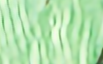}\\
    & Original  & Bicubic       & SRCNN~\cite{dong2016SRCNNPAMI}             &  FSRCNN~\cite{dong2016FSRCNN}        & LapSRN~\cite{lai2017LapSRN}\\
    &PSNR/SSIM  & 24.69/0.7873  & 26.26/0.8487      & 26.38/0.8500  & 26.92/0.8752 \\

    &
    \includegraphics[width=.145\textwidth,valign=t]{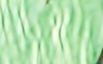}&
    \includegraphics[width=.145\textwidth,valign=t]{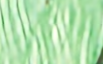}&  
    \includegraphics[width=.145\textwidth,valign=t]{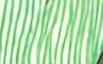}&    \includegraphics[width=.145\textwidth,valign=t]{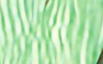}&    
    \includegraphics[width=.145\textwidth,valign=t]{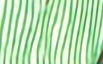}\\
    MANGA109 (4$\times$) &VDSR~\cite{kim2016VDSR}          &DRRN~\cite{tai2017DRRN}           & RCAN~\cite{zhang2018RCAN}         & CARN~\cite{ahn2018CARN}   & Ours\\
    YumeiroCooking       &26.92/0.8731  &27.20/0.8822   & 29.85/0.9368 & 27.58/0.8953   & \textbf{30.33/0.9422}\\
    
\end{tabular}
\end{center}
\vspace*{-4mm}
\caption{\textbf{Visual comparison for 4$\times$.} Super-resolution comparison on sample images with sharp edges and texture, taken from URBAN100~\cite{huang2015URBAN100} and MANGA109~\cite{fujimoto2016manga109} for the scale of 4$\times$. The sharpness of the edges on the objects and textures restored by our method is the best.}
\label{fig:BI_4x}
\vspace*{-4mm}
\end{figure*}
\vspace*{-5mm}
\subsection{Implementation}
In this section of the paper, we present the implementation details of our system. In each cascading residual on the residual block, we have three ($k$=3) DRLMs, and in each DRLM, we have three RBs densely connected, one compression unit and one Laplacian attention. The filter size in all the layers is set to 3$\times$3 with dilation of 1$\times$1 except in the Laplacian pyramid where it is three, five and seven. Likewise, the number of feature maps in all the convolutional layers are fixed to 64, except the last reconstruction layer where the output is either one for grayscale or three for color images. To keep the size of the feature maps the same, zeros are padded accordingly. In pyramid attention, the feature maps are reduced by a factor of four. We also use a post-upscaling procedure instead of pre-scaling for more efficient processing and to avoid the pre-scaling artifacts.

\section{Experiments}
In this section, we first examine the contributions of various elements of our proposed network. Then we test the model on five publicly available super-resolution dataset, namely,   SET5~\cite{bevilacqua2012Set5}, SET14~\cite{zeyde2010Set14}, URBAN100~\cite{huang2015URBAN100}, B100~\cite{martin2001BSD100}, and MANGA109~\cite{fujimoto2016manga109}. The metrics employed for evaluation are PSNR and SSIM on the luminance channel obtained through YCbCr color space. We also give a comparison on object recognition performance against the competing super-resolution methods.

\subsection{Training settings} To make fair comparisons with the current state-of-the-art use CNN methods~\cite{wang2018esrgan,zhang2018RCAN,lim2017EDSR}, we use the same settings which are specified in their particular papers. Similar to~\cite{wang2018esrgan}, we train our network on DIV2K and Flicker2K datasets~\cite{timofte2017ntireFlicker2K}. Furthermore, we diversify the training images through data augmentation, which is accomplished by random rotations using multiples of 90$\degree$ supplemented via horizontal and vertical flipping. The batch size is 16 while the size of the low-resolution input is $48 \times 48$. To optimize the system, ADAM~\cite{kingma2014adam} is utilized with the default parameters of $\beta_1$ =0.9, $\beta_2$ =0.999, and $\epsilon= 10^{-8}$. The learning rate is fixed to $10^{-4}$ originally and then decreased to half after every  $2 \times 10^5$ iterations. The network is designed utilizing the PyTorch framework~\cite{paszke2017pytorch} on a Tesla P100 GPU.

\subsection{Ablation Studies}
\subsubsection{Influence of the skip connections}
Skip connections are the backbone of the current state of the art network~\cite{kim2016VDSR,ahn2018CARN,zhang2018RDN,zhang2018RCAN}. Here, we demonstrate the effectiveness of the skip-connections \ie Long skip connection (LSC), Medium skip connection (MSC), and dense local connections (DLC), in our model. The connections are categorized based on their length. Table~\ref{table:Skip_connections} shows the average PSNR on SET5 for 2$\times$ for the different settings of connections. The PSNR is higher when all the connections are present while the performance relatively downgrades when some of the connections are absent.  In the absence of all connections, the depth of the network does not yield benefit. This experiment illustrates the significance of the different connection types for our deep network.

\begin{figure*}
\begin{center}
\begin{tabular}{c@{ } c@{ }  c@{ } c@{ } c@{ } c}
    
    \multirow{4}{*}{\includegraphics[width=.24\textwidth,height=5.3cm,valign=t]{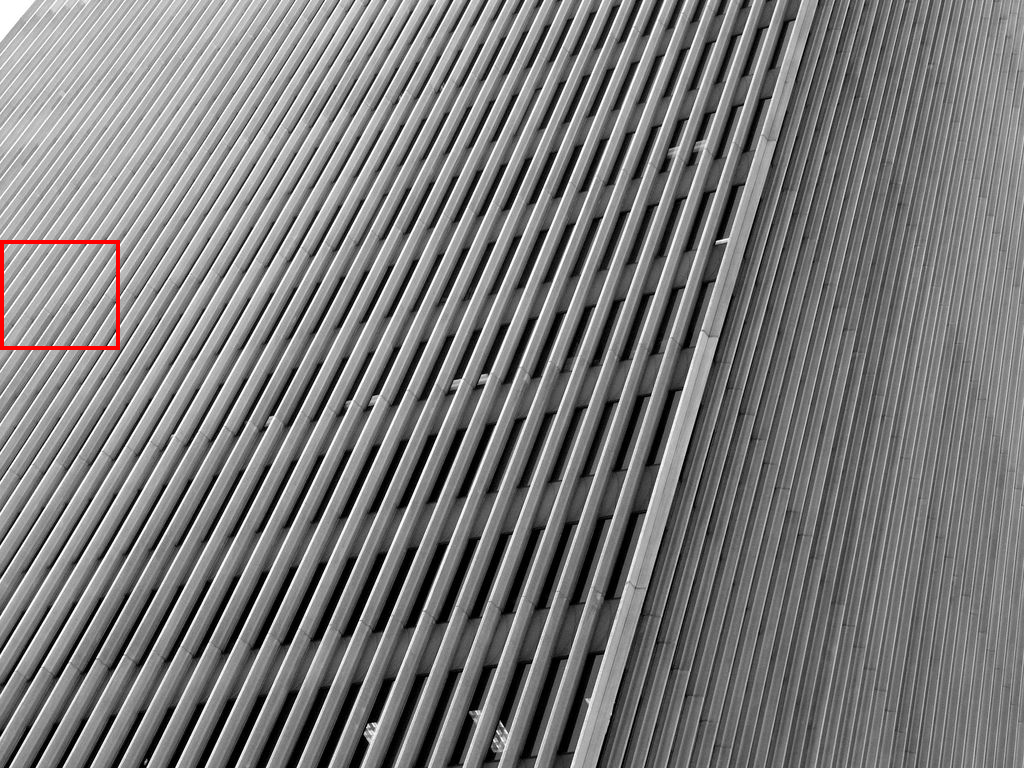}} &  
    \includegraphics[width=.145\textwidth,trim={0cm 1cm 0cm 1cm},clip,valign=t]{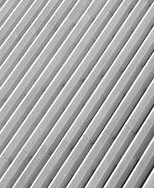}&
  	\includegraphics[width=.145\textwidth,trim={0cm 1cm 0cm 1cm},clip,valign=t]{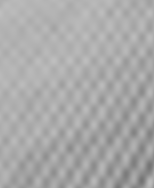}&   
    \includegraphics[width=.145\textwidth,trim={0cm 1cm 0cm 1cm},clip,valign=t]{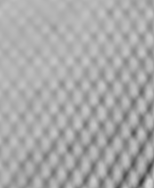}&   
  	\includegraphics[width=.145\textwidth,trim={0cm 1cm 0cm 1cm},clip,valign=t]{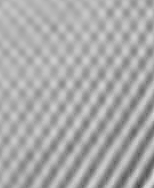}&
  	\includegraphics[width=.145\textwidth,trim={0cm 1cm 0cm 1cm},clip,valign=t]{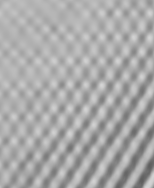}\\
    & Original  & Bicubic       & SRCNN~\cite{dong2016SRCNNPAMI}         &  FSRCNN~\cite{dong2016FSRCNN}         & VDSR~\cite{kim2016VDSR}\\
    &PSNR/SSIM  & 18.41/0.3189  & 18.79/0.3416  & 19.01/0.3519  & 19.10/0.3578 \\

    &
    \includegraphics[width=.145\textwidth,trim={0cm 1cm 0cm 1cm},clip,valign=t]{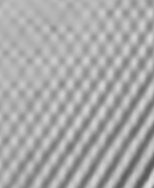}&
    \includegraphics[width=.145\textwidth,trim={0cm 1cm 0cm 1cm},clip,valign=t]{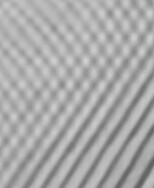}&  
    \includegraphics[width=.145\textwidth,trim={0cm 1cm 0cm 1cm},clip,valign=t]{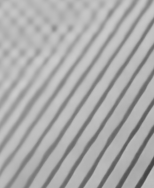}&    
    \includegraphics[width=.145\textwidth,trim={0cm 1cm 0cm 1cm},clip,valign=t]{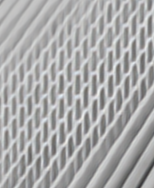}&  
    \includegraphics[width=.145\textwidth,trim={0cm 1cm 0cm 1cm},clip,valign=t]{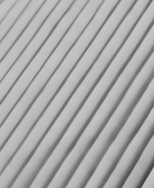}\\
    Urban100 (8$\times$) & DRCN~\cite{kim2016DRCN}           & DRRN~\cite{tai2017DRRN}          &MSLapSRN~\cite{MSLapSRN}      & RCAN~\cite{zhang2018RCAN}          & Ours\\
    img$\_$045           & 19.18/0.3728   & 19.76/0.4043  & 18.87/0.3819 & 19.93/0.4035  & \textbf{20.58/0.4549} \\

    \multirow{4}{*}{\includegraphics[width=.24\textwidth,height=2.95cm,valign=t]{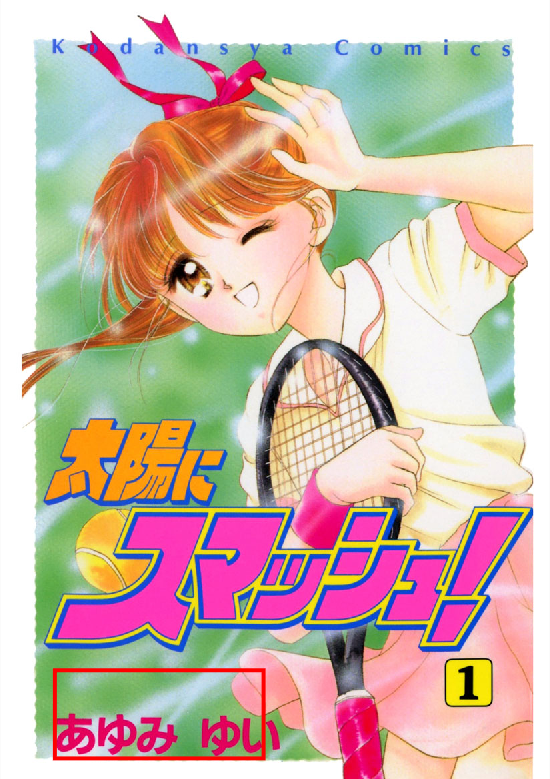}} &  
    \includegraphics[width=.145\textwidth,valign=t]{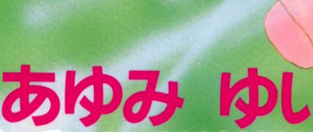}&
  	\includegraphics[width=.145\textwidth,valign=t]{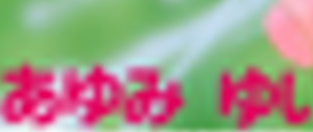}&   
    \includegraphics[width=.145\textwidth,valign=t]{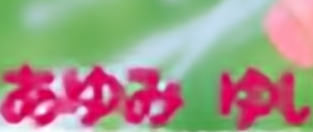}&   
 	\includegraphics[width=.145\textwidth,valign=t]{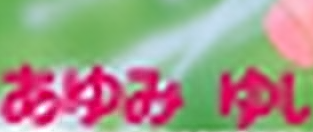}&
  	\includegraphics[width=.145\textwidth,valign=t]{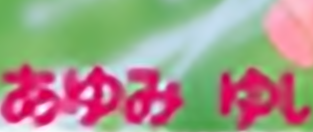}\\
    & Original  & Bicubic       & SRCNN~\cite{dong2016SRCNNPAMI}         &  FSRCNN~\cite{dong2016FSRCNN}         & VDSR~\cite{kim2016VDSR}\\
    &PSNR/SSIM  & 24.91/0.7578  & 25.97/0.7894  & 26.07/0.7766  & 26.37/0.7981 \\

    &
    \includegraphics[width=.145\textwidth,valign=t]{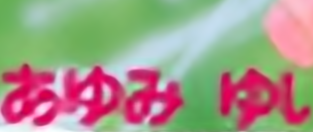}&
    \includegraphics[width=.145\textwidth,valign=t]{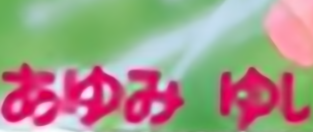}&  
    \includegraphics[width=.145\textwidth,valign=t]{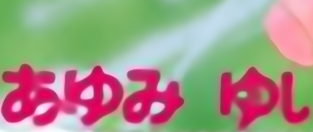}&    
    \includegraphics[width=.145\textwidth,valign=t]{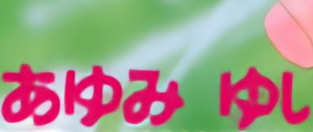}&
    \includegraphics[width=.145\textwidth,valign=t]{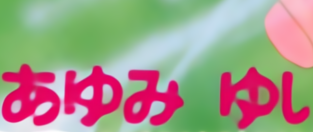}\\
    MANGA109~\cite{fujimoto2016manga109} (8$\times$) & DRCN~\cite{kim2016DRCN}         & DRRN~\cite{tai2017DRRN}          &MSLapSRN~\cite{MSLapSRN}      & RCAN~\cite{zhang2018RCAN}          & Ours\\
    TaiyouNiSmash       &27.09/0.8216  &27.60/0.8378   &28.02/0.8532  & 30.68/0.8962 &\textbf{31.09/0.9032}\\
\end{tabular}
\end{center}
\vspace*{-3mm}
\caption{\textbf{Visual comparison for 8$\times$.} Comparisons on images with fine details for a high upsampling factor of 8$\times$ on URBAN100~\cite{huang2015URBAN100} and MANGA109~\cite{fujimoto2016manga109}. The best results are in bold.}
\label{fig:BI_8x}
\vspace*{-5mm}
\end{figure*}

\subsubsection{Laplacian attention} 
The second essential component of our model is Laplacian attention. We provide a comparison of the network with and without the use of Laplacian attention in Table \ref{table:Skip_connections}. The results shown support our claim that the selection of essential features through multiple frequency bands assist enhancement of the image and improve the overall accuracy. It should be considered that super-resolution techniques \cite{lim2017EDSR,haris2018DDBPN} have matured greatly since SRCNN~\cite{dong2014SRCNN} and further improvement requires sophisticated network design and the weighting of features through appropriate selection criteria. Both of the mentioned provisions are achieved in our model through Laplacian pyramid attention, and cascading with residual on the residual architecture.

\subsubsection{Parameters, runtime, and depth analysis} 
The number of parameters is a crucial factor in determining the potential of the CNN networks. More parameters usually lead to better performance, however, computing more parameters requires deeper networks, which increases the computational load. Our aim here is to utilize previously computed features; hence, concatenating the earlier computed features maps strike a balance between performance, depth, and run time. Furthermore, our network achieves state-of-the-art performance with a mere 160 convolutional layers as compared to RCAN~\cite{zhang2018RCAN} \ie 400+. In Figure~\ref{fig:PvsP}, we provide a comparison of the parameters and the performance. Our model has fewer parameters as compared to EDSR~\cite{lim2017EDSR} and the runtime is less as compared to RCAN~\cite{zhang2018RCAN} \ie the time taken by RCAN for an image of size 824$\times$1168 on average is 1.14s opposed to our method 0.045s on MANGA109~\cite{huang2015URBAN100} for 4$\times$. This efficiency is due to the fact that our method mainly uses concatenations instead of expensive addition operations. In Figure~\ref{fig:PvsT}, the runtime comparisons are provided against state-of-the-art methods. The PSNR and efficiency are higher for our model, which essentially demonstrates that compact models can push the boundaries with non-conventional architectures.

\subsection{Comparisons}
We present the comparison of our model with the state-of-the-art CNN models which include SRCNN~\cite{dong2014SRCNN}, FSRCNN~\cite{dong2016FSRCNN}, VDSR~\cite{kim2016VDSR}, SCN~\cite{wang2015SCN}, SPMSR~\cite{peleg2014SPMSR}, LapSRN~\cite{lai2017LapSRN}, MSLapSRN~\cite{MSLapSRN}, MemNet~\cite{tai2017memnet}, EDSR~\cite{lim2017EDSR}, SRMDNF~\cite{zhang2018SRMDNF}, D-DBPN~\cite{haris2018DDBPN}, IRCNN~\cite{zhang2017IrCNN}, RDN~\cite{zhang2018RDN}, RCAN~\cite{zhang2018RCAN} and CARN~\cite{ahn2018CARN}. Similar to~\cite{lim2017EDSR}, we employ self-ensemble to boost the performance of our model and denote it with a '+' to differentiate it from the single model. 

Similar to contemporary state-of-the-art models, we experiment on two types of image degradations; bicubic-downsample and blur-downsample~\cite{zhang2017IrCNN}. For evaluation of the models, the bicubic downsampling scales of 2$\times$, 3$\times$, 4$\times$, and 8$\times$ are adopted, while blur-downsampling is achieved through a Gaussian kernel having 1.6$\sigma^2$ for a scale of 3$\times$.  

\begin{table*}[!t]
\caption{\textbf{Quantitative evaluation of competing methods.} We report the performance of state-of-the-art algorithms on widely used publicly available datasets, in terms of PSNR (in dB) and SSIM. The best results are highlighted with \textcolor{red}{red} color while the \textcolor{blue}{blue} color represents the second best SR.}
\vspace*{-3mm}
\begin{center}
\begin{tabular}{l|c|cc||cc||cc||cc||cc}
\hline
Method &Scale &\multicolumn{2}{c||} {SET5~\cite{bevilacqua2012Set5}}
& \multicolumn{2}{c||} {SET14~\cite{zeyde2010Set14}}
& \multicolumn{2}{c||} {BSD100~\cite{martin2001BSD100}}
& \multicolumn{2}{c||} {URBAN100~\cite{huang2015URBAN100}}
& \multicolumn{2}{c} {MANGA109~\cite{fujimoto2016manga109}}\\ \hline \hline
         &   &PSNR  &SSIM   &PSNR  &SSIM   &PSNR  &SSIM   &PSNR  &SSIM   &PSNR  &SSIM\\ \hline \hline
Bicubic                          & &33.66 &0.9299 &30.24 &0.8688 &29.56 &0.8431 &26.88 &0.8403 &30.80 &0.9339\\
SRCNN~\cite{dong2016SRCNNPAMI}   & &36.66 &0.9542 &32.45 &0.9067 &31.36 &0.8879 &29.50 &0.8946 &35.60 &0.9663\\
 FSRCNN~\cite{dong2016FSRCNN}    & &37.05 &0.9560 &32.66 &0.9090 &31.53 &0.8920 &29.88 &0.9020 &36.67 &0.9710\\
VDSR~\cite{kim2016VDSR}          & &37.53 &0.9590 &33.05 &0.9130 &31.90 &0.8960 &30.77 &0.9140 &37.22 &0.9750\\
LapSRN~\cite{lai2017LapSRN}      & &37.52 &0.9591 &33.08 &0.9130 &31.08 &0.8950 &30.41 &0.9101 &37.27 &0.9740\\
MemNet~\cite{tai2017memnet}      & &37.78 &0.9597 &33.28 &0.9142 &32.08 &0.8978 &31.31 &0.9195 &37.72 &0.9740\\
EDSR~\cite{lim2017EDSR}          & &38.11 &0.9602 &33.92 &0.9195 &32.32 &0.9013 &32.93 &0.9351 &39.10 &0.9773\\
SRMDNF~\cite{zhang2018SRMDNF}    &2$\times$ &37.79 &0.9601 &33.32 &0.9159 &32.05 &0.8985 &31.33 &0.9204 &38.07 &0.9761\\
D-DBPN~\cite{haris2018DDBPN}     & &38.09 &0.9600 &33.85 &0.9190 &32.27 &0.9000 &32.55 &0.9324 &38.89 &0.9775\\
RDN~\cite{zhang2018RDN}          & &38.24 &0.9614 &34.01 &0.9212 &32.34 &0.9017 &32.89 &0.9353 &39.18 &0.9780\\
RCAN~\cite{zhang2018RCAN}        & &\textcolor{blue}{38.27} &0.9614 &34.12 &0.9216 &32.41 &0.9027 &33.34 &0.9384 &39.44 &0.9786\\
CARN~\cite{ahn2018CARN}          & &37.76  &0.9590 &33.52 &0.9166 &32.09 &0.8978   &31.92 &0.9256  &38.36 &0.9764\\
DRLN (ours)    & & \textcolor{blue}{38.27} & \textcolor{blue}{0.9616} & \textcolor{blue}{34.28} & \textcolor{blue}{0.9231} & \textcolor{blue}{32.44} & \textcolor{blue}{0.9028} & \textcolor{blue}{33.37} & \textcolor{blue}{0.9390} & \textcolor{blue}{39.58} & \textcolor{blue}{0.9786}\\
DRLN+ (ours)                   & &\textcolor{red}{38.34} &\textcolor{red}{0.9619} &\textcolor{red}{34.43} &\textcolor{red}{0.9247} &\textcolor{red}{32.47} &\textcolor{red}{0.9032} &\textcolor{red}{33.54} &\textcolor{red}{0.9402} &\textcolor{red}{39.75} &\textcolor{red}{0.9792}\\
\hline \hline

Bicubic                          & &30.39 &0.8682 &27.55 &0.7742 &27.21 &0.7385 &24.46 &0.7349 &26.95 &0.8556\\
SRCNN~\cite{dong2016SRCNNPAMI}   & &32.75 &0.9090 &29.30 &0.8215 &28.41 &0.7863 &26.24 &0.7989 &30.48 &0.9117\\
 FSRCNN~\cite{dong2016FSRCNN}   & &33.18 &0.9140 &29.37 &0.8240 &28.53 &0.7910 &26.43 &0.8080 &31.10 &0.9210\\
VDSR~\cite{kim2016VDSR}        & &33.67 &0.9210 &29.78 &0.8320 &28.83 &0.7990 &27.14 &0.8290 &32.01 &0.9340\\
LapSRN~\cite{lai2017LapSRN}    & &33.82 &0.9227 &29.87 &0.8320 &28.82 &0.7980 &27.07 &0.8280 &32.21 &0.9350\\
MemNet~\cite{tai2017memnet}   & &34.09 &0.9248 &30.00 &0.8350 &28.96 &0.8001 &27.56 &0.8376 &32.51 &0.9369\\
EDSR~\cite{lim2017EDSR}       &3$\times$ &34.65 &0.9280 &30.52 &0.8462 &29.25 &0.8093 &28.80 &0.8653 &34.17 &0.9476\\
SRMDNF~\cite{zhang2018SRMDNF}  & &34.12 &0.9254 &30.04 &0.8382 &28.97 &0.8025 &27.57 &0.8398 &33.00 &0.9403\\
RDN~\cite{zhang2018RDN}     & &34.71 &0.9296 &30.57 &0.8468 &29.26 &0.8093 &28.80 &0.8653 &34.13 &0.9484\\
RCAN~\cite{zhang2018RCAN}    & &34.74 &0.9299 &30.65 &0.8482 &29.32 &0.8111 &29.09 &0.8702 &34.44 &0.9499\\
CARN~\cite{ahn2018CARN}   & &34.29&0.9255 &30.29&0.8407 &29.06&0.8034 &28.06&0.8493 &33.49& 0.9440\\
DRLN (ours)    & &\textcolor{blue}{34.78} &\textcolor{blue}{0.9303} &\textcolor{blue}{30.73} &\textcolor{blue}{0.8488} &\textcolor{blue}{29.36} &\textcolor{blue}{0.8117} &\textcolor{blue}{29.21} &\textcolor{blue}{0.8722} &\textcolor{blue}{34.71} &\textcolor{blue}{0.9509}\\
DRLN+ (ours)   & &\textcolor{red}{34.86} &\textcolor{red}{0.9307} &\textcolor{red}{30.80} &\textcolor{red}{0.8498} &\textcolor{red}{29.40} &\textcolor{red}{0.8125} &\textcolor{red}{29.37} &\textcolor{red}{0.8746} &\textcolor{red}{34.94} &\textcolor{red}{0.9518}\\ 
\hline \hline

Bicubic & &28.42 &0.8104 &26.00 &0.7027 &25.96 &0.6675 &23.14 &0.6577 &24.89 &0.7866\\
SRCNN~\cite{dong2016SRCNNPAMI}   & &30.48 &0.8628 &27.50 &0.7513 &26.90 &0.7101 &24.52 &0.7221 &27.58 &0.8555\\
 FSRCNN~\cite{dong2016FSRCNN}  & &30.72 &0.8660 &27.61 &0.7550 &26.98 &0.7150 &24.62 &0.7280 &27.90 &0.8610\\
VDSR~\cite{kim2016VDSR}    & &31.35 &0.8830 &28.02 &0.7680 &27.29 &0.0726 &25.18 &0.7540 &28.83 &0.8870\\
LapSRN~\cite{lai2017LapSRN}  & &31.54 &0.8850 &28.19 &0.7720 &27.32 &0.7270 &25.21 &0.7560 &29.09 &0.8900\\
MemNet~\cite{tai2017memnet}  & &31.74 &0.8893 &28.26 &0.7723 &27.40 &0.7281 &25.50 &0.7630 &29.42 &0.8942\\
EDSR~\cite{lim2017EDSR}    & &32.46 &0.8968 &28.80 &0.7876 &27.71 &0.7420 &26.64 &0.8033 &31.02 &0.9148\\
SRMDNF~\cite{zhang2018SRMDNF}  &4$\times$ &31.96 &0.8925 &28.35 &0.7787 &27.49 &0.7337 &25.68 &0.7731 &30.09 &0.9024\\
D-DBPN~\cite{haris2018DDBPN}  & &32.47 &0.8980 &28.82 &0.7860 &27.72 &0.7400 &26.38 &0.7946 &30.91 &0.9137\\
RDN~\cite{zhang2018RDN}     & &32.47 &0.8990 &28.81 &0.7871 &27.72 &0.7419 &26.61 &0.8028 &31.00 &0.9151\\
RCAN~\cite{zhang2018RCAN}    & &32.63 &0.9002 &28.87 &0.7889 &27.77 &0.7436 &26.82 &0.8087 &31.22 &0.9173\\
CARN~\cite{ahn2018CARN}    & &32.13 &0.8937 &28.60   &0.7806 &27.58  &0.7349 &26.07  &0.7837 & 30.40  & 0.9082\\
DRLN (ours)    & &\textcolor{blue}{32.63} &\textcolor{blue}{0.9002} &\textcolor{blue}{28.94} &\textcolor{blue}{0.7900} &\textcolor{blue}{27.83} &\textcolor{blue}{0.7444} &\textcolor{blue}{26.98} &\textcolor{blue}{0.8119} &\textcolor{blue}{31.54} &\textcolor{blue}{0.9196}\\ 
DRLN+ (ours)   & &\textcolor{red}{32.74} &\textcolor{red}{0.9013} &\textcolor{red}{29.02} &\textcolor{red}{0.7914} &\textcolor{red}{27.87} &\textcolor{red}{0.7453} &\textcolor{red}{27.14} &\textcolor{red}{0.8149} &\textcolor{red}{31.78} &\textcolor{red}{0.9211}\\ \hline \hline

Bicubic & &24.40 &0.6580 &23.10 &0.5660 &23.67 &0.5480 &20.74 &0.5160 &21.47 &0.6500\\
SRCNN~\cite{dong2016SRCNNPAMI}   & &25.33 &0.6900 &23.76 &0.5910 &24.13 &0.5660 &21.29 &0.5440 &22.46 &0.6950\\
 FSRCNN~\cite{dong2016FSRCNN}  & &20.13 &0.5520 &19.75 &0.4820 &24.21 &0.5680 &21.32 &0.5380 &22.39 &0.6730\\
SCN~\cite{wang2015SCN}     & &25.59 &0.7071 &24.02 &0.6028 &24.30 &0.5698 &21.52 &0.5571 &22.68 &0.6963\\
VDSR~\cite{kim2016VDSR}     & &25.93 &0.7240 &24.26 &0.6140 &24.49 &0.5830 &21.70 &0.5710 &23.16 &0.7250\\
LapSRN~\cite{lai2017LapSRN}  & &26.15 &0.7380 &24.35 &0.6200 &24.54 &0.5860 &21.81 &0.5810 &23.39 &0.7350\\
MemNet~\cite{tai2017memnet}  &8$\times$ &26.16 &0.7414 &24.38 &0.6199 &24.58 &0.5842 &21.89 &0.5825 &23.56 &0.7387\\
MSLapSRN~\cite{MSLapSRN}   & &26.34 &0.7558 &24.57 &0.6273 &24.65 &0.5895 &22.06 &0.5963 &23.90 &0.7564\\
EDSR~\cite{lim2017EDSR}    & &26.96 &0.7762 &24.91 &0.6420 &24.81 &0.5985 &22.51 &0.6221 &24.69 &0.7841\\
D-DBPN~\cite{haris2018DDBPN}  & &27.21 &0.7840 &25.13 &0.6480 &24.88 &0.6010 &22.73 &0.6312 &25.14 &0.7987\\
RCAN~\cite{zhang2018RCAN}    & &27.31 &0.7878 &25.23 &0.6511 &24.98 &0.6058 &23.00 &0.6452 &25.24 &0.8029\\
DRLN (ours) & & \textcolor{blue}{27.36} & \textcolor{blue}{0.7882} & \textcolor{blue}{25.34} & \textcolor{blue}{0.6531} & \textcolor{blue}{25.01} & \textcolor{blue}{0.6057} & \textcolor{blue}{23.06} & \textcolor{blue}{0.6471} & \textcolor{blue}{25.29} & \textcolor{blue}{0.8041}\\
DRLN+ (ours) & & \textcolor{red}{27.46} & \textcolor{red}{0.7916} & \textcolor{red}{25.40} & \textcolor{red}{0.6547} & \textcolor{red}{25.06} & \textcolor{red}{0.6070} & \textcolor{red}{23.24} & \textcolor{red}{0.6523} & \textcolor{red}{25.55} & \textcolor{red}{0.8087}
\\ \hline \hline
\end{tabular}
\label{table:bicubic}
\end{center}
\vspace*{-5mm}
\end{table*}

\subsubsection{Bicubic degradations} 
In this section, we provide the qualitative results of our model against competitive methods in Figure~\ref{fig:BI_4x} and~\ref{fig:BI_8x}. 

\noindent 
\textbf{4$\times$ Visual Comparisons:} 
To be fair in comparison, the images furnished for qualitative comparison in Figure~\ref{fig:BI_4x} are from the same dataset images as RCAN~\cite{zhang2018RCAN}. The first two images are from URBAN100~\cite{huang2015URBAN100}, and the last picture is from MANGA109~\cite{fujimoto2016manga109} for upscaling of 4$\times$. As shown in Figure~\ref{fig:BI_4x}, all competing methods, in general, fail to recover edges and introduce blurring artifacts. In \enquote{img\_076}, the competing CNN algorithms are unable to retrieve the rectangular shapes and blur out the edges and boundaries representing the outlines of the windows. Our method is faithful to the original image, providing results with proper rectangular structures and straight lines. 

Similarly, in the second example, \ie \enquote{img\_044} in Figure~\ref{fig:BI_4x}, most of the methods distort the horizontal lines and blur out the background. Furthermore, the orientation of the lines on the cropped parts is in the opposite direction and forms a checkerboard pattern for~\cite{dong2016SRCNNPAMI,kim2016VDSR}. Moreover, a close inspection reveals that RCAN~\cite{zhang2018RCAN} and CARN~\cite{ahn2018CARN} fused the background with the horizontal lines, removing the sharpness from the images shown while in our case, the lines are clearly visible and separated from the background, recovering sharp details. 

The last image in Figure~\ref{fig:BI_4x} is from MANGA109~\cite{fujimoto2016manga109} for 4$\times$ scale. The textures in the cropped regions are blemished and mixed by together most of the state-of-the-art methods and are unable to recover the shape of the green colored strokes except for RCAN~\cite{zhang2018RCAN}. However, RCAN~\cite{zhang2018RCAN} is also unequipped to super-resolve the lines correctly as it can be witnessed that the lines are blended in the right side of the cropped image. Furthermore, the super-resolved green lines are blurry in the case of RCAN~\cite{zhang2018RCAN}. Our network can super-resolve most of the details and textures without producing visible artifacts from the shown low-resolution image. The green lines are sharp and closer in structure to the original.

\begin{figure*}
\begin{center}
\begin{tabular}{c@{ } c@{ }  c@{ } c@{ } c@{ } c}

    \multirow{4}{*}{\includegraphics[width=.24\textwidth,trim={0cm 0.2cm 0 0},clip,valign=t]{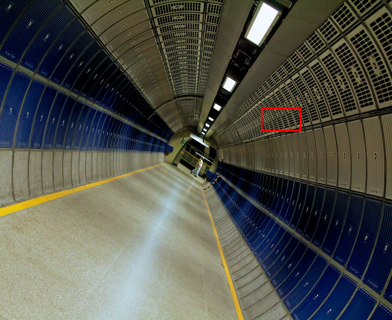}} &  
    \includegraphics[width=.145\textwidth,valign=t]{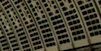}&
    \includegraphics[width=.145\textwidth,valign=t]{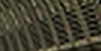}&   
	\includegraphics[width=.145\textwidth,valign=t]{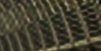}&    
  	\includegraphics[width=.145\textwidth,valign=t]{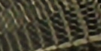}&
  	  	\includegraphics[width=.145\textwidth,valign=t]{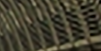}\\
    & Original& Bicubic &SRCNN~\cite{dong2016SRCNNPAMI} & FSRCNN~\cite{dong2016FSRCNN}  & VDSR~\cite{kim2016VDSR}  \\
    &PSNR/SSIM& 26.10/0.7032  &27.91/0.7874 &24.34/0.6711&28.34/0.8166 \\

    &
    \includegraphics[width=.145\textwidth,valign=t]{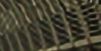}&
    \includegraphics[width=.145\textwidth,valign=t]{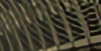}&    
  	\includegraphics[width=.145\textwidth,valign=t]{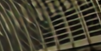}&
    \includegraphics[width=.145\textwidth,valign=t]{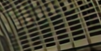}&
    \includegraphics[width=.145\textwidth,trim={0.5cm 0.1cm 0 0},clip,valign=t]{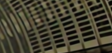}\\
    URBAN100 (3$\times$)   &IRCNN~\cite{zhang2017IrCNN} &SRMDNF~\cite{zhang2018SRMDNF} &RDN~\cite{zhang2018RDN} &RCAN~\cite{zhang2018RCAN} & Ours\\
    img\_078 & 28.57/0.8184 &29.08/0.8342 &29.94/0.8513 &30.65/0.8624& \textbf{31.13/0.8685}\\

    \multirow{8}{*}{\includegraphics[width=.24\textwidth,trim={0cm 4cm 0 1.5cm},clip,valign=t]{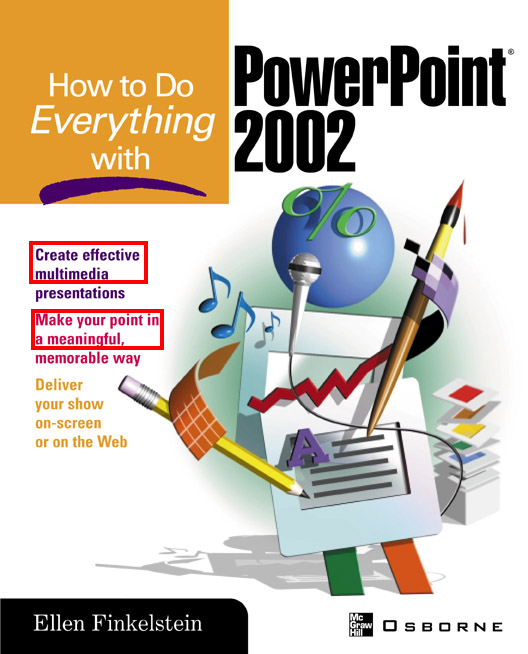}} &  
    \includegraphics[width=.145\textwidth,valign=t]{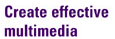}&
    \includegraphics[width=.145\textwidth,valign=t]{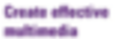}&
  	\includegraphics[width=.145\textwidth,valign=t]{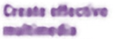}&
	\includegraphics[width=.145\textwidth,valign=t]{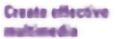}&    
  	\includegraphics[width=.145\textwidth,valign=t]{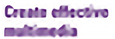}\\
  	 &
    \includegraphics[width=.145\textwidth,valign=t]{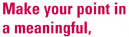}&
    \includegraphics[width=.145\textwidth,valign=t]{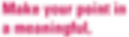}&
  	\includegraphics[width=.145\textwidth,valign=t]{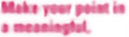}&
	\includegraphics[width=.145\textwidth,valign=t]{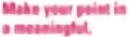}&    
  	\includegraphics[width=.145\textwidth,valign=t]{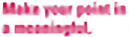}\\
    & Original& Bicubic      & FSRCNN~\cite{dong2016FSRCNN}        &VDSR~\cite{kim2016VDSR}  & IRCNN~\cite{zhang2017IrCNN}\\
    &PSNR/SSIM& 22.58/0.84597 &21.09/0.8254 &19.30/0.6960 &24.60/0.9092 \\

    &
    \includegraphics[width=.145\textwidth,valign=t]{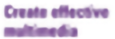}&
    \includegraphics[width=.145\textwidth,valign=t]{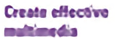}&
  	\includegraphics[width=.145\textwidth,valign=t]{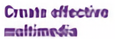}&
	\includegraphics[width=.145\textwidth,valign=t]{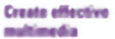}&    
  	\includegraphics[width=.145\textwidth,valign=t]{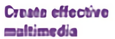}\\
  	 &
    \includegraphics[width=.145\textwidth,valign=t]{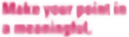}&
    \includegraphics[width=.145\textwidth,valign=t]{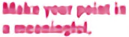}&
  	\includegraphics[width=.145\textwidth,valign=t]{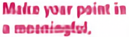}&
	\includegraphics[width=.145\textwidth,valign=t]{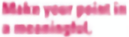}&    
  	\includegraphics[width=.145\textwidth,valign=t]{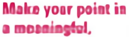}\\
    URBAN100 (3$\times$) &EDSR~\cite{lim2017EDSR}        &SRMDNF~\cite{zhang2018SRMDNF}       &RCAN~\cite{zhang2018RCAN}          &CARN~\cite{ahn2018CARN} & Ours\\
    img\_062             &24.48/0.9105 &28.63/0.9695 &29.41/0.9775 &24.45/0.9096& \textbf{30.78/0.9830}\\

\end{tabular}
\end{center}
\vspace*{-3mm}
\caption{\textbf{Blur-Downscale (BD) degradation.} Comparison on sample images with sharp edges and texture, taken from URBAN100 and SET14 datasets for the scale of 3$\times$. The sharpness of the edges on the objects and textures restored by our method is the best.}
\label{fig:BD_3x}
\vspace*{-3mm}
\end{figure*}

\begin{table*}
\caption{\textbf{Quantitative results with blur-down degradation.} The best results are highlighted with \textcolor{red}{red} color while the \textcolor{blue}{blue} color represents the second best.}
\vspace*{-3mm}
\begin{center}
\begin{tabular}{l|c||cc||cc||cc||cc||cc}\hline
&&\multicolumn{2}{c||} {SET5~\cite{bevilacqua2012Set5}}
& \multicolumn{2}{c||} {SET14~\cite{zeyde2010Set14}}
& \multicolumn{2}{c||} {BSD100~\cite{martin2001BSD100}}
& \multicolumn{2}{c||} {URBAN100~\cite{huang2015URBAN100}}
& \multicolumn{2}{c} {MANGA109~\cite{fujimoto2016manga109}}\\ \cline{3-12}
Method      &Scale      & PSNR  & SSIM  & PSNR  & SSIM  & PSNR  & SSIM  & PSNR  & SSIM  & PSNR  & SSIM  \\  \hline\hline
Bicubic     &           &28.78  &0.8308 &26.38  &0.7271 &26.33  &0.6918 &23.52  &0.6862 &25.46  &0.8149\\
SPMSR       &           &32.21  &0.9001 &28.89  &0.8105 &28.13  &0.7740 &25.84  &0.7856 &29.64  &0.9003\\
SRCNN~\cite{dong2016SRCNNPAMI}       &           &32.05  &0.8944 &28.80  &0.8074 &28.13  &0.7736 &25.70  &0.7770 &29.47  &0.8924\\
 FSRCNN~\cite{dong2016FSRCNN}      &           &26.23  &0.8124 &24.44  &0.7106 &24.86  &0.6832 &22.04  &0.6745 &23.04  &0.7927\\
VDSR~\cite{kim2016VDSR}        &           &33.25  &0.9150 &29.46  &0.8244 &28.57  &0.7893 &26.61  &0.8136 &31.06  &0.9234\\
IRCNN~\cite{zhang2017IrCNN}       & 3$\times$ &33.38  &0.9182 &29.63  &0.8281 &28.65  &0.7922 &26.77  &0.8154 &31.15  &0.9245\\
SRMDNF~\cite{zhang2018SRMDNF}      &           &34.01  &0.9242 &30.11  &0.8364 &28.98  &0.8009 &27.50  &0.8370 &32.97  &0.9391\\
RDN~\cite{zhang2018RDN}         &           &34.58  &0.9280 &30.53  &0.8447 &29.23  &0.8079 &28.46  &0.8582 &33.97  &0.9465\\
RCAN~\cite{zhang2018RCAN}        &           &34.70  &0.9288 &30.63  &0.8462 &29.32  &0.8093 &28.81  &0.8647 &34.38  &0.9483\\
DRLN (ours)         &           &\textcolor{blue}{34.81} & \textcolor{blue}{0.9297} & \textcolor{blue}{30.81} & \textcolor{blue}{0.8487} & \textcolor{blue}{29.40} & \textcolor{blue}{0.8121} & \textcolor{blue}{29.11} & \textcolor{blue}{0.8697} & \textcolor{blue}{34.84} & \textcolor{blue}{0.9506}\\
DRLN+ (ours)        &           & \textcolor{red}{34.87} & \textcolor{red}{0.9301} & \textcolor{red}{30.86} & \textcolor{red}{0.8495} & \textcolor{red}{29.44} & \textcolor{red}{0.8128} & \textcolor{red}{29.26} & \textcolor{red}{0.8718} & \textcolor{red}{35.07} & \textcolor{red}{0.9516}\\\hline 
\end{tabular}
\end{center}
\label{table:BD_3x}
\vspace*{-5mm}
\end{table*}

\noindent
\textbf{8$\times$ Visual Comparisons:}
To show the powerful reconstruction ability of our network, we present extreme examples of 8$\times$ super-resolution in Figure~\ref{fig:BI_8x} from URBAN100~\cite{huang2015URBAN100} and MANGA109~\cite{fujimoto2016manga109}. Because of the significant scaling factor in image \enquote{img\_045}, the models~\cite{dong2014SRCNN,kim2016VDSR,kim2016DRCN} using bicubic upsampled input creates artificial checkerboard artifacts and structures in the images due to the incorrect initial input. While, on the other hand, the recent state-of-the-art method~\cite{zhang2018RCAN} which takes the low-resolution photographs as input, is unable to recover the high frequencies reliably due to the notable upscaling and also produces new structures instead of recovering the original lines. MSLapSRN~\cite{MSLapSRN} is able to super-resolve the lines correctly in the lower half of the image, and this may be due to the progressive reconstruction \ie employing loss after every 2$\times$ resolution to achieve 8$\times$ output. In our case, the lines are super-resolved correctly without employing multiple losses. This shows the super-resolution capability of our CNN model.  

\begin{figure*}
\begin{center}
\begin{tabular}{c@{}c c@{}c c@{}c c@{}c}
   \multicolumn{2}{c} {\includegraphics[width=.24\textwidth,valign=t]{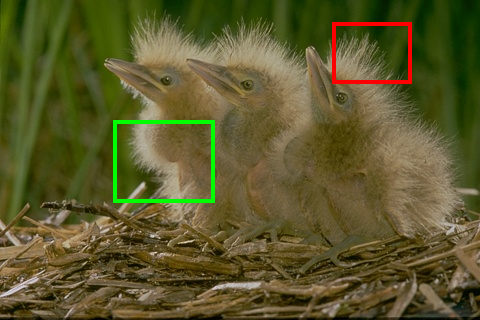}}&
   \multicolumn{2}{c} {\includegraphics[width=.24\textwidth,valign=t]{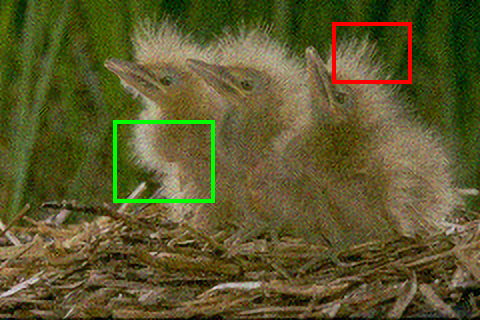}}&
   \multicolumn{2}{c} {\includegraphics[width=.24\textwidth,valign=t]{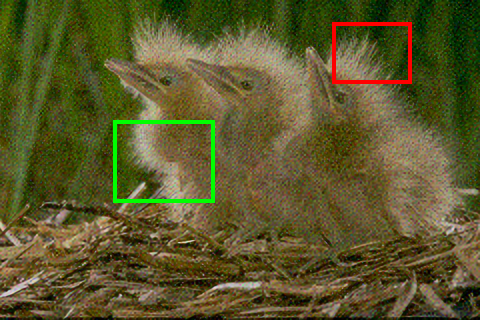}}&
   \multicolumn{2}{c} {\includegraphics[width=.24\textwidth,valign=t]{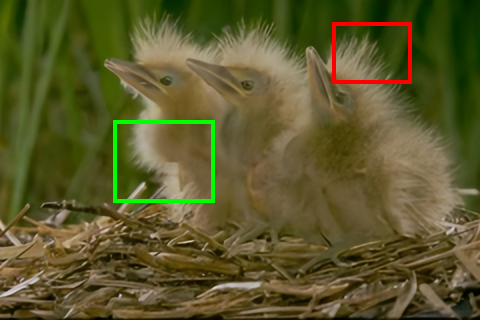}}\\
    \vspace*{-3mm}

  &&&&&&&\\
    \includegraphics[width=.115\textwidth,valign=t]{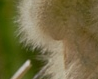}&
    \includegraphics[width=.123\textwidth,valign=t]{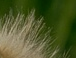}&
    
    \includegraphics[width=.115\textwidth,valign=t]{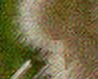}&
    \includegraphics[width=.123\textwidth,valign=t]{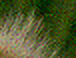}&

    \includegraphics[width=.115\textwidth,valign=t]{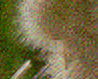}&
    \includegraphics[width=.123\textwidth,valign=t]{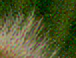}&
    
    \includegraphics[width=.115\textwidth,valign=t]{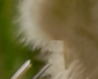}&
    \includegraphics[width=.123\textwidth,valign=t]{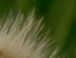}\\

    \multicolumn{2}{c}{Original(PSNR/SSIM)}&  \multicolumn{2}{c}{IRCNN~\cite{zhang2017IrCNN}~(28.73/0.6762)}  &  \multicolumn{2}{c}{RCAN~\cite{zhang2018RCAN}~(28.44/0.6607)} &  \multicolumn{2}{c}{Ours~(\textbf{32.46/0.8760})} \\

\end{tabular}
\end{center}
\vspace*{-3mm}
\caption{\textbf{Noisy SR visual Comparison on BSD100.} Textures on the birds are much better reconstructed, and the noise removed by our method as compared to the IRCNN~\cite{zhang2017IrCNN} and RCAN~\cite{zhang2018RCAN} for $\sigma=10$.}
\label{fig:noisy_bsd2x}
\vspace*{-0.5mm}

\begin{center}
\begin{tabular}{c@{ } c@{ } c@{ } c@{ } c}
\includegraphics[width=0.1\textwidth]{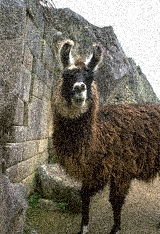}&
\includegraphics[width=0.2\textwidth]{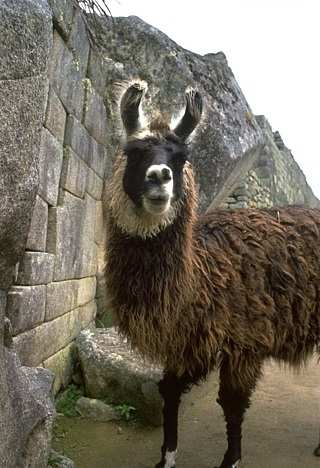}&
\includegraphics[width=0.2\textwidth]{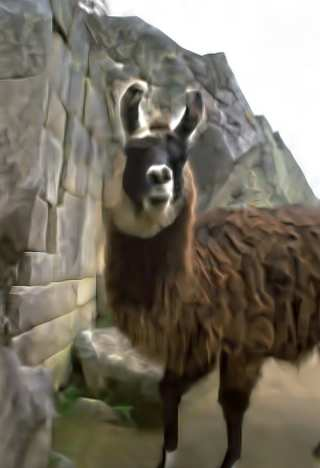}&
\includegraphics[width=0.2\textwidth]{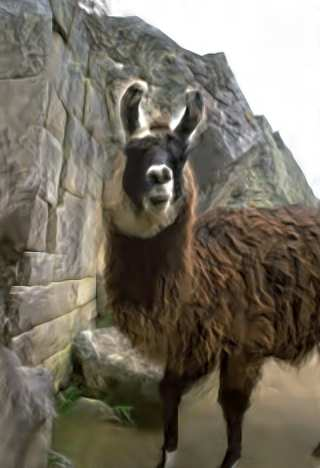}&
\includegraphics[width=0.2\textwidth]{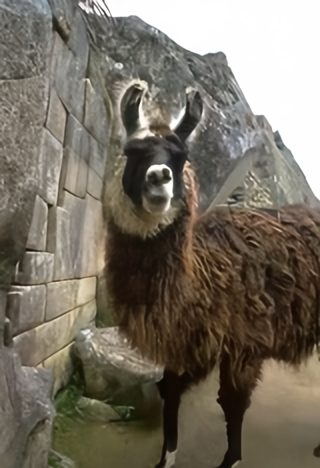}\\
Noisy           & GT        & BM3D-SR~\cite{Dabov2007BM3D}       &  BM3D-SRNI~\cite{singh2014SRNI}    & Ours\\
$\sigma = 20$   & PSNR/SSIM & 25.05/0.5868  & 25.31/0.6206 & \textbf{27.03/0.7330}\\
\end{tabular}
\end{center}
\vspace*{-3mm}
\caption{\textbf{Noisy visual comparison on Llama.} Textures on the fur, and on rocks in the background are much better reconstructed in our result as compared to the conventional BM3D-SR and BM3D-SRNI. }
\label{fig:lama_image}
\vspace*{-5mm}
\end{figure*}

The second low-resolution image, in Figure~\ref{fig:BI_8x} for 8$\times$ is from MANGA109~\cite{fujimoto2016manga109}, titled, \enquote{TaiyouNiSmash}, contains minimal high-frequencies and hence is challenging to super-resolve to the desired outcome. Nevertheless, our method still can reproduce better results and avoids producing blurring and artificial structures as compared to competitive techniques. The algorithms of VDSR~\cite{kim2016VDSR}, MSLapSRN~\cite{MSLapSRN}, \etc creates blurry outputs and unsharp images. Similarly, RCAN~\cite{zhang2018RCAN} can produce slightly sharp edges than its predecessor; however, it connected the gaps in different letters. Our model is more faithful to the original image and better captures the small gaps present in the letters.

\noindent
\textbf{Quantitative Comparisons:} Table~\ref{table:bicubic} shows the quantitative results for all the competing methods. These results are borrowed from their corresponding published papers. Our scheme outperforms all other approaches on all datasets for all scales which complements the visual sequences presented earlier in Figures~\ref{fig:BI_4x} and~\ref{fig:BI_8x}. Our model quantitative results outperform even without employing the self-ensemble technique.

The improvement of our method on SET5~\cite{bevilacqua2012Set5} and SET14~\cite{zeyde2010Set14} is marginal due to a small number of images (the number with the dataset shows the images present \ie SET5 has only five photographs, and SET14 has only 14 images) in these mentioned datasets; however, a clear trend emerges when the number of images increases. For example, on MANGA109~\cite{fujimoto2016manga109}, the average PSNR increment across all scales for our model is 0.34dB and 3.98dB compared to second leading method \ie RCAN~\cite{zhang2018RCAN} and the pioneering SRCNN~\cite{dong2014SRCNN}, respectively.

\begin{figure}[t]
\begin{center}
\begin{tabular}{c@{ } c}
\includegraphics[width=0.48\textwidth]{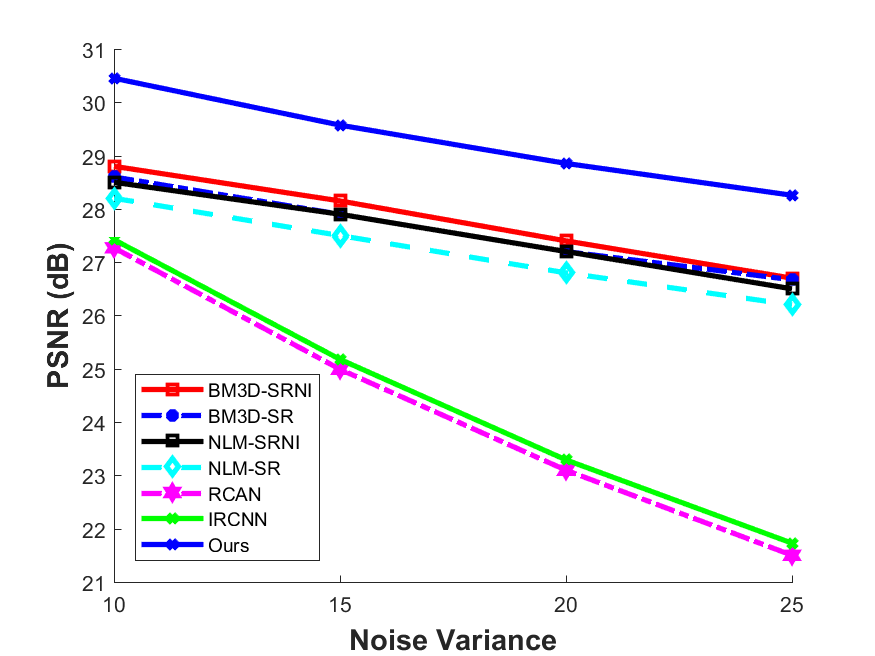}&
\end{tabular}
\end{center}
\vspace*{-3mm}
\caption{\textbf{Noisy super-resolution.} The plots show average PSNR as functions of noise sigma. Our method consistently improves over specific noisy super-resolution methods and CNN for all $\sigma$ levels.}
\label{fig:Noisy_graphs}
\vspace*{-7mm}
\end{figure}

\subsubsection{Blur-Downscale (BD) degradations} 

More recently, blur-down (BD) degraded images~\cite{zhang2017IrCNN} are super-resolved to showcase the potential of super-resolution architectures. We also utilize blur-downsampled photographs to compare against the state-of-the-art. 

 \noindent 
\textbf{3$\times$ Visual Comparisons:} We first present three examples; two from URBAN100~\cite{huang2015URBAN100} and one from MANGA109~\cite{fujimoto2016manga109} in Fig.~\ref{fig:BD_3x}. The images from URBAN100~\cite{huang2015URBAN100} super-resolved by the competing methods contain blur effects near the upper end of the buildings as can be seen in the cropped versions. The only methods which perform comparatively better are RDN~\cite{zhang2018RDN} and RCAN~\cite{zhang2018RCAN}; however, our approach is not only able to remove the blur but also restore the high-frequency details. This may be due to the Laplacian-attention used at the end of each DRLM, which captures the important discriminative features that are useful for high-frequency restoration. 

Next, we evaluate our algorithm on an image from the classical SET14~\cite{zeyde2010Set14} shown in Figure~\ref{fig:BD_3x}. In the crop sections, our method produces relatively sharp edges and crisper text than state-of-the-art algorithms which mostly exhibit blurry and distorted text. IRCNN~\cite{zhang2017IrCNN}, SRMD~\cite{zhang2018SRMDNF}, and RCAN~\cite{zhang2018RCAN} are specifically designed to handle blur-downscale super-resolution; however, our method produces best qualitative results having more than 1dB PSNR for this particular image.

\noindent 
\textbf{Quantitative Comparisons:} Next, Table \ref{table:BD_3x} provides the comparison against nine competitive methods. Here, again RDN~\cite{zhang2018RDN} and RCAN~\cite{zhang2018RCAN} shows good results compared to~\cite{dong2014SRCNN,zhang2017IrCNN,kim2016VDSR}; however, our single and self-ensemble models achieve a notable performance gain over all methods in general, and RDN~\cite{zhang2018RDN} and RCAN~\cite{zhang2018RCAN} particular. The average PSNR gain over both the mentioned methods for 3$\times$ super-resolution of blur-down degraded images is 0.55dB and 0.33dB for all the datasets. Our architecture better generalizes the task at hand,  and our Laplacian attention can select the relevant features more reliably in contrast to RCAN's~\cite{zhang2018RCAN} channel attention.

\begin{table*}[t]
\caption{\textbf{Objection recognition.} We report the performance of ResNet~\cite{he2016ResNet} using different SR algorithms as a pre-processing step.}
\vspace*{-5mm}
\begin{center}
\begin{tabular}{l|cccccccc}
Evaluation  &Bicubic &DRCN~\cite{kim2016DRCN}  & FSRCNN~\cite{dong2016FSRCNN} &PSyCo~\cite{perez2016psyco}& ENet-E~\cite{sajjadi2017enhancenet}& RCAN~\cite{zhang2018RCAN}  & Ours & Baseline \\ \hline \hline
Top-1 error &0.506   &0.477 &0.437  &0.454& 0.449 & 0.393 & 0.345 &0.260\\
Top-5 error &0.266   &0.242 &0.196  &0.224& 0.214 & 0.167 & 0.121 &0.072\\
\end{tabular}
\end{center}
\label{table:Object_recog_error}
\vspace*{-6mm}
\end{table*}

\subsubsection{Noisy downscale (ND) degradations} 

\noindent 
\textbf{2$\times$ Visual Comparisons:} 
We present two images for super-resolving the noisy images. Figure~\ref{fig:noisy_bsd2x} shows the comparison of our method with CNN methods on the Birds image from BSD100~\cite{martin2001BSD100} for a low-level noise ($\sigma=10$). It can be observed that our method significantly recovered more texture and removed the noise close to the ground truth image. IRCNN~\cite{zhang2017IrCNN} and RCAN~\cite{zhang2018RCAN} fail to remove the noise, rather they amplify it. The other noisy image, \enquote{Llama}, shown in Figure~\ref{fig:lama_image} is taken from Singh~\etal~\cite{singh2014SRNI} for a fair comparison against traditional algorithms. The difference in texture details on the fur and the background can be observed in our case and the conventional methods. Our method can super-resolve the fur more appropriately as compared to the other methods. 

\noindent 
\textbf{Quantitative Comparisons:} 
To compare quantitatively, we follow the footsteps of Singh~\etal~\cite{singh2014SRNI} which uses the first 50 images from the BSD100~\cite{martin2001BSD100}. We compare against super-resolving noisy image algorithms, which include (SRNI)~\cite{singh2014SRNI}, BM3D-SR~\cite{Dabov2007BM3D} and NLM-SR~\cite{Buades2005NLM} as well as the image restoration algorithm IRCNN~\cite{zhang2017IrCNN}. Moreover, BM3D-SR or NLM-SR indicates applying the traditional denoising approach first \ie (BM3D or NLM), followed by image super-resolution (SR). 

Currently, RCAN~\cite{zhang2018RCAN} is state-of-the-art in single image super-resolution; however, we show that it is unable to handle noisy images. In Figure~\ref{fig:Noisy_graphs}, we present the quantitative results with the competing algorithms for four noise levels \ie $\sigma~=~10,15,20$ and $25$. Our algorithm constantly outperforms CNN-based algorithms and specifically designed methods at all noise levels. Furthermore, the CNN methods~\cite{zhang2017IrCNN,zhang2018RCAN} performance is relative to the traditional algorithms when the noise levels are low \ie $\sigma~=~10$ and $15$; however, it degrades significantly as $\sigma$ increases while, on the other hand, the performance of our algorithm is better at high-noise levels as well.

\begin{figure}
\begin{center}
\begin{tabular}{c@{ } c@{ }  c}

    \multirow{3}{*}{\includegraphics[width=.225\textwidth,trim={0cm 0.0cm 2.2cm 0},clip,valign=t]{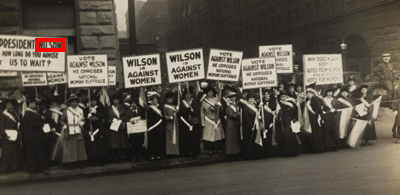}} &  
    \includegraphics[width=.12\textwidth,valign=t]{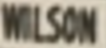}&
    \includegraphics[width=.12\textwidth,valign=t]{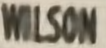}\\
    & Bicubic & SRCNN~\cite{dong2016SRCNNPAMI}  \\

    &
    \includegraphics[width=.12\textwidth,valign=t]{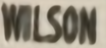}&
    \includegraphics[width=.12\textwidth,valign=t]{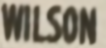}\\
    LR input   & DRCN~\cite{kim2016DRCN} &Ours\\
    
    \multirow{3}{*}{\includegraphics[width=.225\textwidth,trim={0cm 0.0cm 0cm 0.45cm},clip,valign=t]{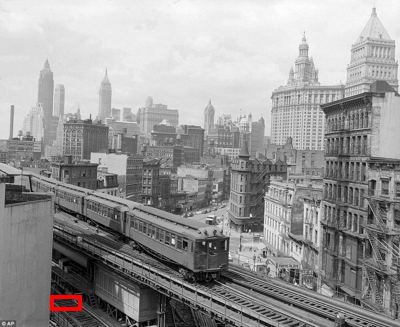}} &  
    \includegraphics[width=.12\textwidth,valign=t]{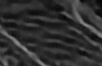}&
    \includegraphics[width=.12\textwidth,valign=t]{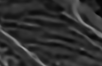}\\
    & FSRCNN~\cite{dong2016FSRCNN} & VDSR~\cite{kim2016VDSR}  \\
    
        &
    \includegraphics[width=.12\textwidth,valign=t]{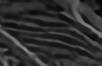}&
    \includegraphics[width=.12\textwidth,valign=t]{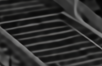}\\
    LR input  & MSLapSRN~\cite{MSLapSRN} &Ours\\

\end{tabular}
\end{center}
\vspace*{-3mm}
\caption{\textbf{Comparison of real-world images.} In these cases, neither the downsampling blur kernels nor the ground-truth images are available. In the top photograph, our methods reconstruct the letters \enquote{W} and \enquote{I} correctly while the competing methods combine the letters. Similarly, in the bottom picture, the rail is accurately super-resolved without any artifacts while others fail to super-resolve the horizontal lines correctly.}
\label{fig:Historic_4x}
\vspace*{-3mm}
\end{figure}

\subsubsection{Real-World historic super-resolution}

In this section, we illustrate the application of our algorithm on real-world historic~\cite{lai2017LapSRN} images suffering from JPEG compression artifacts. The information about the downsampling operators and ground-truth images are unavailable. The results of our reconstruction against state-of-the-art algorithms are shown in Figure~\ref{fig:Historic_4x}. The output of our algorithm is clear and sharper as shown in both Figures.

\subsection{Performance on Object Recognition}

As discussed earlier, image restoration tasks assist in high-level computer vision tasks; therefore, we demonstrate and analyze the effectiveness of our method on object recognition current state-of-the-art super-resolution techniques. We use the same settings as~\cite{zhang2018RCAN}, which evaluates the performance on the initial 1k images from ImageNet~\cite{deng2009imagenet}. The image of size 224$\times$224 is downscaled by 4$\times$ to achieve the image size of 56$\times$56. The downscaled versions are then upscaled to the original size images via the super-resolution algorithms and subsequently fed through ResNet50~\cite{he2016deep} for classifications. The accuracy of the classification network is used to determine the potential of the super-resolution algorithms. 

We compare with six methods \ie Bicubic, DRCN~\cite{kim2016DRCN}, FSRCNN~\cite{dong2016FSRCNN}, PSyCo~\cite{perez2016psyco}, ENet-E~\cite{sajjadi2017enhancenet} and RCAN~\cite{zhang2018RCAN}. The top-1 and top-5 errors for object recognition are recorded in Table \ref{table:Object_recog_error}. Our method is more accurate as it provides the lowest error; hence, this illustrates the ability of our network to reconstruct appropriate frequencies more reliably.    

\subsection{Limitations}
\begin{figure}
\begin{center}
\begin{tabular}{c@{ } c@{ } c@{ } c}

    \multirow{3}{*}{\includegraphics[width=.14\textwidth,trim={0cm 0.0cm 0cm 0},clip,valign=t]{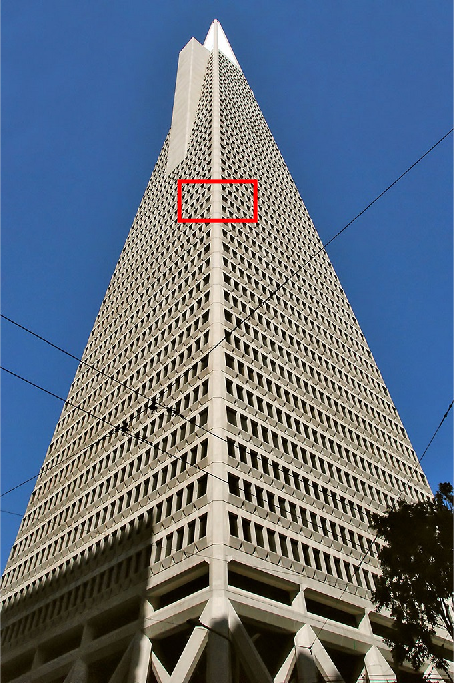}} &  
    \includegraphics[width=.105\textwidth,valign=t]{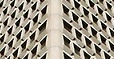}&
    \includegraphics[width=.105\textwidth,valign=t]{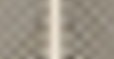}&
    \includegraphics[width=.105\textwidth,valign=t]{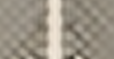}\\
    & GT & Bicubic & SRCNN~\cite{dong2016SRCNNPAMI} \\

    &
    \includegraphics[width=.105\textwidth,valign=t]{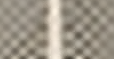}&
    \includegraphics[width=.105\textwidth,valign=t]{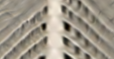}&
    \includegraphics[width=.105\textwidth,valign=t]{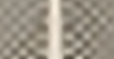}\\
    & FSRCNN~\cite{dong2016FSRCNN}  & SelfExSR~\cite{huang2015SelfEx} &  DRCN~\cite{kim2016DRCN}\\
    
        &
   \includegraphics[width=.105\textwidth,valign=t]{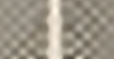}&
    \includegraphics[width=.105\textwidth,valign=t]{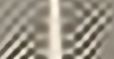}&
    \includegraphics[width=.105\textwidth,valign=t]{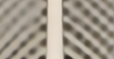}\\
    Ground-truth& VDSR~\cite{kim2016VDSR}& LapSRN~\cite{lai2017LapSRN} &Ours\\

\end{tabular}
\end{center}
\vspace*{-3mm}
\caption{\textbf{Limitation.} A failure case for super-resolution of 8$\times$. Our algorithm is not able to create finer details if the input low-resolution images lack sufficient high-frequency details.}
\label{fig:limitation_x8}
\vspace*{-3mm}
\end{figure}

Our model has shown the ability to render sharp and clean images for all upsampling scales; however, it struggles to \enquote{hallucinate} finer details. For example, an image with 8$\times$ is shown in Figure~\ref{fig:limitation_x8}, the top of the building is very challenging as due to the large downsampling operator \ie 8$\times$. All the algorithms fail to recover the fine details at this level. 
The traditional methods, \eg SelfExSR~\cite{huang2015SelfEx}, which exploit the self-similarity and 3D scene geometry, also fail to recover fine details. Similarly, MS-LapSRN~\cite{MSLapSRN} progressively upsamples to produce the 8$\times$ results as opposed to ours where the 8$\times$ upsampling is achieved directly; however,~\cite{MSLapSRN} is unable to give the desired outcome. Furthermore, this limitation is common to all the super-resolution methods~\cite{MSLapSRN,zhang2018RCAN,dong2016SRCNNPAMI,kim2016VDSR}.

\section{Conclusion}
In this exposition, we propose a modular convolution neural network for highly accurate image super-resolution. We also employ various components to boost the performance of super-resolution. We thoroughly analyze and present a comprehensive evaluation of the choice of our network design. 

We employ cascading residual on the residual structure to design a large depth network using long skip connection, short skip connection, and local connections. The cascading residual on the residual architecture helps in the flow of low-frequency information to make network learn high and mid-level frequency information. We use densely connected residual blocks, which re-use the previously computed features. This type of setting has multiple advantages such as implicit \enquote{deep supervision} and learning from high-level complex features. We also introduce Laplacian attention, which models the essential features on multiple scales and learns the inter and intra-level dependencies between the feature maps. 

Furthermore, we perform an extensive evaluation of super-resolution datasets, low-resolution noisy images and real-world images (unknown blur downsampling). We also have shown the results on Bicubic and blur-down kernels to demonstrate the effectiveness of our proposed methods. Further, we present the performance of object recognition on the super-resolved images by different methods. We have illustrated the potential of our network for image super-resolution; however, our network is general and can be applied to other low-level vision tasks such as image restoration, synthesis, and transformation problems.  
\vspace*{-2mm}
\ifCLASSOPTIONcaptionsoff
  \newpage
\fi

\bibliographystyle{IEEEtran}
\bibliography{ref}

\end{document}